\documentclass[12pt]{article}
\usepackage{setspace}
\usepackage{amsmath}
\usepackage{amssymb}
\usepackage[top=1in,bottom=1in,right=0.5in,left=0.5in]{geometry}
\usepackage{caption}
\usepackage{psfrag}
\usepackage{graphicx}
\usepackage{hyperref}
\usepackage{subfigure}
\usepackage{authblk}
\usepackage{verbatim}
\usepackage{float}
\usepackage{bibcheck}
\usepackage[backend=biber,
            style=numeric,
            sorting=none,
            giveninits=true,
            maxnames=5,
            doi=false,
            isbn=false,
            url=false]{biblatex}
\title{Non-singular Brane/Bubble Cosmologies \\ in the Presence of String Cloud}
\author{Karma P. Sherpa\thanks{E-Mail: sherpa.karma.pincho@gmail.com and karma\_202310028@smit.smu.edu.in} }
\author{Rishi Pokhrel\thanks{E-Mail: rishipokhrel.smit@gmail.com and rishi\_20211037@smit.smu.edu.in} }                     
\author{Tanay K. Dey\thanks{E-mail: tanay.dey@gmail.com and tanay.d@smit.smu.edu.in}}
\affil{Department of Physics, Sikkim Manipal Institute of Technology, Sikkim Manipal University, Majitar, Sikkim-737136, India.}

\date{}
\begin{document}
	\maketitle

\begin{abstract}\noindent
 We investigate the cosmological evolution of a four-dimensional brane/bubble universe embedded in a five-dimensional Anti-de Sitter bulk containing a uniformly distributed cloud of strings. The endpoints of the strings attached to the brane induce an effective matter component in the four-dimensional universe. We derive the Friedmann equations for both the Randall-Sundrum braneworld and the shellworld (dark bubble) scenarios and show that one can obtain a nonsingular bounce in both the brane and bubble universes under certain conditions of the black hole mass and the four-dimensional cosmological constant. Though the bounce due to the spatial curvature occurs outside the black hole horizon, however, the bounce obtained due to the presence of string cloud in the braneworld scenario suffers from the Cauchy horizon instability as the bounce occurs inside the Cauchy horizon of the black hole. In contrast, for the shellworld scenario, the bounce can be made to occur outside the black hole horizon, thereby avoiding the horizon instability. We also analyze the field fluctuations on the brane/bubble and show that the fluctuations remain finite and well-behaved at the bounce point.
\end{abstract}

\clearpage


\section{Introduction}\noindent

The standard cosmological model based on general relativity successfully explains a wide range of cosmological observations yet faces significant challenges. One of the most profound issues is the presence of an initial singularity. At the point of singularity, the energy density and the spacetime curvature diverges, leading to an inconsistency in the classical description of general relativity. This suggests the necessity of alternative approaches, either by modifying the gravitational theory itself or by altering the fundamental nature of spacetime.

Motivated by the Anti de Sitter (AdS)/ Conformal Field Theory (CFT) correspondence \cite{Maldacena:1997re}, braneworld models \cite{Randall:1999vf,Randall:1999ee} were introduced where our universe is interpreted as a $(3+1)$-dimensional hypersurface incorporated in a $(4+1)$-dimensional AdS bulk spacetime. The hypersurface divides the AdS geometry in two identical regions on either side of the brane which imposes a $\mathbb{Z}_2$ symmetry. There are two possible approaches to study the cosmological evolution of the brane universe. In the first approach called the Binetruy-Deffayet-Langlos (BDL) approach, the brane is considered to be fixed at a certain position in the bulk and the bulk metric is made time dependent \cite{Kanti:1999sz,Flanagan:1999cu,Binetruy:1999hy}. In the second approach, the bulk metric is kept static and the brane is allowed to move in the radial direction of the bulk \cite{Chamblin:1999ya,Kraus:1999it,Nojiri:2002hz}. The motion of the brane in the bulk results in a time dependent induced metric on the brane which can be interpreted as the cosmological evolution of the brane universe. In \cite{Barcelo2000}, a charge parameter is considered in the bulk AdS-Schwarzschild spacetime and cosmology on the boundary of the spacetime is studied.

A singularity free solution for the charged AdS-Schwarzschild spacetime is obtained in \cite{Mukherji:2002ft}, where the scale factor of the brane smoothly evolves from contraction to expansion. The black hole charge, interpreted as stiff matter on the brane, contributes a negative energy density, which provides nonsingular or bouncing cosmological scenarios. Such bouncing brane universe was further studied through the BDL approach along with the moving domain wall approach in \cite{Myung:2002vg}. The study was generalized to higher dimensional bulk with arbitrary number of dimensions in \cite{Biswas:2003rb}. Bouncing cosmological scenarios were investigated in \cite{Foffa:2003gt} for dilaton-gravity in the braneworld framework. However, it was shown in \cite{Hovdebo:2003ug} that the brane bounces back from inside the Cauchy horizon of the charged black hole and is unstable for small perturbations indicating that the brane encounters a singularity at the transition point. By considering a $SU(2)$ gauge bulk field, it was shown in \cite{Okuyama:2004in} that a nonsingular brane universe free from instabilities can be obtained. Such instabilities can also be circumvented by considering a 3-brane embedded in a warped five-dimensional background containing a dilaton and a Kalb–Ramond two-form field \cite{DeRisi:2007dn}. The bounce in this case is induced by the negative dark radiation term. The nonsingular two dimensional brane universe is studied in a four dimensional black hole spacetime with scalar hair in \cite{Biswas:2005zn}. The nature of bouncing brane universe in the background of Born-Infeld bulk and charged Gauss-Bonnet bulk were studied in \cite{Mukherji:2008hs}, while bouncing universe with a vacuum Gauss-Bonnet bulk was studied in \cite{Maeda:2007cb,Maeda:2011px}.

Furthermore, in an attempt to obtain dark energy from string theory, the dark bubble or the shellword model was introduced in \cite{Banerjee:2018qey}. A vacuum characterized by the higher cosmological constant, $\Lambda_{5(+)}$ can decay non-perturbatively into a vacuum with the lower cosmological constant, $\Lambda_{5(-)}$ via the nucleation of a Brown-Teitelboim instanton \cite{BROWN1988787}. The decay of a nonperturbative AdS vacuum through bubble nucleation to a lower AdS vacuum leads to a spherical brane (bubble) joining the two different AdS vacua. This results a (3+1)-dimensional dS cosmology on the bubble. The bubble shares a similar construction as the Randall-Sundrum braneworld scenario except it lacks the $\mathbb{Z}_2$ symmetry. This results in an inside-outside geometry of the bubble with different AdS curvatures on either side rather than the inside-inside geometry of the braneworld scenario. The shellworld scenario has been further studied in \cite{Banerjee:2019fzz,Banerjee:2020wix,Banerjee:2020wov,Banerjee:2022ree}.

The emergence of an effective matter component on the brane/bubble universe requires the presence of an appropriate matter source in the bulk spacetime. The matter source should redshift as $a^{-3}$ in the Friedmann equations and should linearly scale with the radial coordinate. This can be achieved with the introduction of cloud of strings in the bulk spacetime. The cloud of strings are uniformly distributed and extended in the radial direction of the bulk with their endpoints attached to the brane \cite{Letelier:1979ej, Chakrabortty:2011sp}. These endpoints represent massive particles in four dimensions effectively leading to a matter contribution in the brane universe. The shellworld scenario have effectively studied the implications of these cloud of strings in the construction of the dark bubble \cite{Banerjee:2019fzz, Banerjee:2020wov}. Braneworld cosmology with cloud of strings in the bulk has been considered in \cite{Park:2020jio,Paul:2025gpk} to study the entanglement entropy. 

In the present work, we investigate the cosmological evolution of a four-dimensional brane/bubble universe embedded in a five-dimensional AdS-Schwarzschild bulk containing a uniformly distributed cloud of strings. We first review the bulk geometry and the corresponding black hole solution, showing that the bulk mass parameter can become negative above a critical string density and that the bulk spacetime contains a Cauchy horizon inside the event horizon. We then derive the effective Friedmann equations and study the cosmological evolution in the braneworld scenario. Exact analytical solutions are obtained for the critical brane, while the non-critical case is analyzed qualitatively by interpreting the Friedmann equation as the motion of a classical particle in an effective potential.

For positive or zero bulk mass parameter, the critical brane generally exhibits singular evolution. For a positive brane cosmological constant, two cosmological branches exist within a certain range of brane cosmological constant and string density: one corresponds to a nonsingular bounce driven by the spatial curvature outside the bulk horizon, while the other evolves from expansion to a big crunch singularity. Above a certain value of brane cosmological constant or string density, these branches merge, leaving only monotonically expanding or collapsing singular solutions. Other non-critical geometries also remain singular. For a negative bulk mass parameter, the critical closed universe exhibits nonsingular bouncing and cyclic evolution, whereas the open and flat universes undergo nonsingular eternal expansion. For a positive brane cosmological constant, two nonsingular branches appear: a bouncing and cyclic universe driven by the negative dark radiation term and a gradually expanding universe induced by the spatial curvature. Increasing the brane cosmological constant or string density causes these branches to merge into a single non-singular eternally expanding universe. Open and flat branes with positive brane cosmological constant also exhibit nonsingular expansion, while a negative brane cosmological constant leads to nonsingular bouncing and cyclic evolution for all spatial geometries. Nevertheless, for all cases, the bounce generated by the negative dark radiation term occurs inside the Cauchy horizon of the bulk black hole and is therefore subject to Cauchy horizon instability.

We therefore investigate the cosmological evolution of the shellworld universe using the same framework adopted for the braneworld scenario to explore the stable nonsingular cosmological model. We find that the shellworld model admits a nonsingular bouncing and cyclic universe even for a positive bulk mass parameter in the critical case, with the bounce driven by the negative energy density associated with the dark radiation term. As the brane cosmological constant increases, two distinct bouncing branches emerge: one driven by the negative dark radiation term and the other by the spatial curvature. These branches eventually merge into a single nonsingular eternally expanding universe for sufficiently large values of the brane cosmological constant or string density. We further show that, for an appropriate choice of parameters, the bounce due to the negative dark radiation can occur outside the bulk horizon, resulting in a physically viable and stable nonsingular evolution. A similar behavior is observed for a negative bulk mass parameter, where suitable parameter tuning ensures that the bounce occurs outside the bulk horizons, thereby avoiding horizon instability.

Finally, we examine the evolution of linear test scalar field fluctuations on the brane/shellworld and find that the perturbations remain finite throughout the evolution, including at the bounce. This demonstrates that the nonsingular background remains regular under linear scalar perturbations. Overall, our results establish that the presence of a cloud of strings in the bulk provides a mechanism for realizing nonsingular bouncing cosmologies in both braneworld and shellworld scenarios. However, the shellworld model offers a more robust framework for achieving viable nonsingular evolution, as the bounce can be arranged to occur outside the bulk horizons, avoiding the instabilities associated with Cauchy horizons in braneworld scenario.

The work is organized as follows: In section \ref{sec 2}, we present the background of the AdS Schwarzschild bulk in the presence of cloud of strings. In section \ref{sec 3}, we analyze the nature of cosmological evolution of the braneworld and the shellworld universes. In section \ref{sec 4}, we analyze the field fluctuations on the brane/bubble.
Finally, in section \ref{sec 5}, we summarize our work.


\section{The background}\label{sec 2}\noindent
We consider a five dimensional AdS-Schwarzschild bulk in the presence of uniformly distributed cloud of infinitely long strings. The cloud of strings is extended in the radial direction of the bulk with their endpoints attached on the four dimensional brane placed perpendicular to the radial coordinate of the bulk. The brane divides the AdS spacetime in two different regions. The end points represent massive particles in the four dimensional theory living on the brane. The action for this configuration can be written as, 
\begin{equation}\label{action}
I = -\frac{1}{2\kappa^2}\int\limits_{M} d^5x\sqrt{-g}\left(R_5-2\Lambda_5\right)-\frac{I_{M}}{\kappa^2}+\frac{1}{\kappa^2}\int\limits_{\partial M}d^4x\sqrt{-\gamma} (K +\sigma).
\end{equation}
Here, $R_5$ represents the Ricci scalar of the five dimensional spacetime, $g_{\mu\nu}$ is the five dimensional bulk metric, $\kappa^2=8\pi G_5$ is the bulk gravitational constant, $l$ is the curvature radius of the AdS spacetime and is related to the bulk cosmological constant as $\Lambda_5 =-\frac{6}{l^2}$. The metric $\gamma_{cd}$ is the induced metric on the brane with tension $\sigma$ and $K$ is the trace of the extrinsic curvature $K_{cd}$ on the brane. The contribution of the cloud of strings is expressed as,
\begin{equation} 
I_{M} =\frac{1}{2}\sum_i \mathcal{T}_{i} \int d^2\xi \sqrt{-h}h^{\alpha \beta} \partial_{\alpha}X^{\mu}\partial_{\beta} X^{\nu} g_{\mu\nu}.
\end{equation}
Here, $h^{\alpha\beta}$ denotes the induced metric on the string world-sheet parameterized by $\xi$, with the $i^{th}$ string having tension $\mathcal{T}_i$. Variation of the action (\ref{action}) with respect to the bulk metric $g_{\mu\nu}$ gives us the bulk equation of motion as,
\begin{equation}\label{EFE}
R_{\mu\nu} -\frac{1}{2}Rg_{\mu\nu} +\Lambda g_{\mu\nu} = T_{\mu\nu},
\end{equation}
where, $T_{\mu\nu}$ represents the stress energy tensor of the string cloud and is given by,
\begin{equation}
T^{\mu\nu}=-\sum_i \mathcal{T}_i \int d^2 \xi \frac{1}{\sqrt{\left|g_{\mu\nu}\right|}} \sqrt{\left|h_{\alpha \beta}\right|} h^{\alpha \beta} \partial_\alpha X^\mu \partial_\beta X^\nu \delta_i^{5}(x-X).
\end{equation}
By considering the strings to have uniform tension $\mathcal T$, the string cloud density can be expressed as \cite{Chakrabortty:2011sp},
\begin{equation}
b(x)=\mathcal T\sum_i \delta^3_i(x-X_i).
\end{equation}
Under the static gauge $t=\xi^0 \;{\rm and}\; r=\xi^1$, the non vanishing components of $T^{\mu\nu}$ can be found as,
\begin{equation}
T^{tt}=-\frac{b g^{tt}}{r^3}, \; T^{rr}= -\frac{b g^{rr}}{r^3}\;{\rm and}\; T^t_t=T^r_r=-\frac{b}{r^3},
\end{equation}
where, $b$ stands for average density of the tension of the strings and can be expressed as, 
\begin{equation} 
\label{alpha}
b = \frac{1}{\omega_3}\int b(x)d^3x={\mathcal T \over \omega_3} \sum_{i=1}^{N} \int \delta_i^{3}(x - X_i) d^3x =  {\mathcal T \over \omega_3} \sum_{i=1}^N 1 =  {{\mathcal T}N \over \omega_3}.
\end{equation}
Here, $\omega_3$ is the volume of a 3-dimensional space, $N$ represents the number of strings and $b$ is considered to be positive. The static and symmetric spacetime metric solution of the equation (\ref{EFE}) can be written by,
\begin{equation}
ds^2_5=-f(r)dt^2+\frac{dr^2}{f(r)}+r^2 \tilde{\gamma}_{ij}dx^idx^j,
\label{bulk_metric}
\end{equation}
with $f(r)$ given by,
\begin{equation}
\label{fr}
f(r)=k+\frac{r^2}{l^2}-\frac{m}{r^{2}}-\frac{2  b}{3 r}.
\end{equation}
Here, $k=0,-1,+1$ represents flat, open, or closed geometries respectively and $\tilde{\gamma}_{ij}$ is the metric for the constant curvature manifold $ M^3$ with the volume $\omega_3=d^3x\sqrt{\gamma}$. The constant $m$ is the mass parameter related to the ADM mass $M$ of the black hole as,
\begin{equation}
M =\frac{3\omega_3m}{2\kappa^2} .
\end{equation}
The metric (\ref{bulk_metric}) admits a black hole solution having a curvature singularity at $r=0$, with the horizon determined by the condition $f(r)=0$. So, the horizon radius $r_h$ satisfies the relation,
\begin{align}
k+\frac{r_h^2}{l^2}-\frac{m}{r_h^{2}}-\frac{2  b}{3 r_h}&=0.\\
{\rm or,}\quad\frac{r_h^4}{l^2}+ kr_h^2 -\frac{2  b r_h}{3}-m &=0.
\end{align}
Since the string density is positive, Descarte’s sign rule indicates that the number of horizon depends only on the nature of mass parameter $m$. As long as the mass parameter is greater or equal to zero, there will be only one horizon called event horizon, otherwise there will be maximum two horizons. The outer horizon corresponds to the event horizon and the inner horizon is called Cauchy horizon.

The mass parameter $m$ of the black hole in terms of the horizon radius can be written as,
\begin{equation}\label{mass}
m = k r_h^2+ \frac{r_h^4}{l^2}-\frac{2  b r_h}{3}.
\end{equation}
\begin{figure}[h]
\centering
\includegraphics[width=0.5\linewidth]{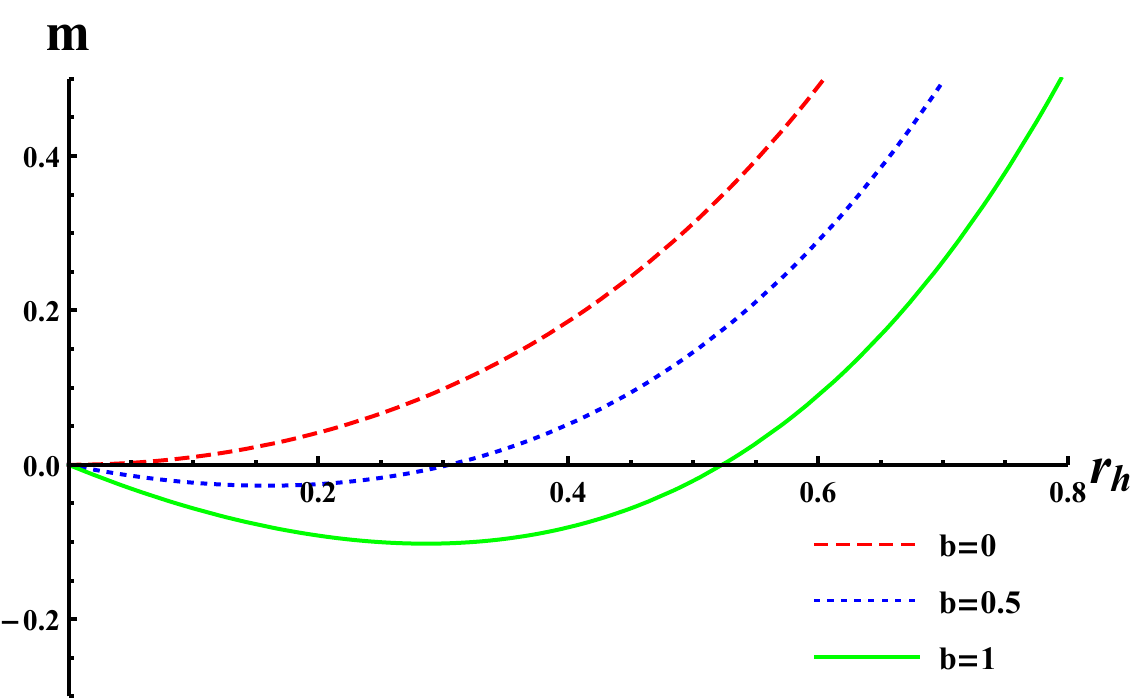}
\caption{Plot of mass as a function of the horizon radius for $k=1$, $l=1$ and various values of string density.}
\label{fig 1}
\end{figure}
From equation (\ref{mass}) and figure (\ref{fig 1}), it is obvious that above a certain value of the string density and for smaller values of the horizon radius, the mass parameter takes a negative value even for $k=+1$. However, even with a negative mass parameter, the entropy of the black hole still maintains one fourth of the area law:
\begin{equation}
S=\frac{r_h^3 \omega_3}{4G_5} .
\end{equation}
The temperature ($T$) of the black hole is expressed as,
\begin{equation} T=\frac{1}{4\pi}\left.\frac{df(r)}{dr}\right|_{r=r_h}=\frac{12r_h^{10}+6kl^2r_h^8-2l^2b r_h^7}{12\pi l^2r_h^9}.
	\end{equation}
With this brief discussion on five dimensional background in the next section, we initiate to study the cosmological evolution of the brane world volume.

	
\section{Cosmological evolution}\label{sec 3}
In order to study the cosmological evolution, we construct a time dependent induced brane metric from $5$-dimensional bulk metric (\ref{bulk_metric}). With this aim, the bulk time $t$ is parameterized by proper time $\tau$ of the braneworld volume as $t=t(\tau)$ which introduces a time-dependent radial coordinate $r=a(\tau)$ in the bulk. For this specific parametrization, we get a 4-dimensional FRW metric on the brane of the form,
\begin{equation}
ds_4^2=-d\tau^2+a^2(\tau)\tilde{\gamma}_{ij}dx^idx^j,
\end{equation}
for the constraint,
\begin{equation}
 -f(a)\left(\frac{dt}{d\tau}\right)^2+\frac{1}{f(a)}\left(\frac{da}{d\tau}\right)^2=-1.
\end{equation}
Therefore, $a(\tau)$ becomes the scale factor as well as position of the brane universe in the radial direction of the bulk. The velocity components of the brane are $v^t= \frac{dt}{d\tau}=\frac{\sqrt{f(a)+\dot{a}^2}}{f(a)}$ and $ v^r=\dot{a}$. The corresponding unit normal vectors are $n^t=\dot{a}/f(a)\;{\rm and}\; n^r=\sqrt{{f(a)+\dot{a}^2}}$. Here, dot ($\dot{}$) denotes differentiation with respect to time $\tau$. The components of the extrinsic curvature of the brane can then be calculated from the relation,  
\begin{equation}
K_{cd}=\frac{1}{2}n^\mu \partial_{\mu} \gamma_{cd}.
\end{equation}
The spatial components of the extrinsic curvature are found to be,
\begin{equation}\label{extrcom}
K_{ij}=\frac{\sqrt{f(a)+\dot{a}^2}}{a}\gamma_{ij},
\end{equation}
where, $i,j$ are the indices of the spatial coordinates on the boundary.

\subsection{The braneworld scenario}
In the Randall-Sundrum braneworld scenario, the brane is considered to be a hypersurface $\partial M$ that separates the bulk spacetime $M$ into two distinct regions. The Israel junction condition \cite{Israel:1966rt} defines the discontinuity of the extrinsic curvature across the brane as,
\begin{equation}	
	[K_{cd}]\equiv K^+_{cd}-K^-_{cd},
\end{equation}
where, $K^+_{cd}$ and $K^-_{cd}$ are the extrinsic curvatures on the two sides of the brane. The junction condition relates the discontinuity of the extrinsic curvature to the energy momentum tensor $S_{cd}$ on the brane as,
\begin{equation}\label{junction}	
	[K_{cd}]=-\kappa^2\left(S_{cd}-\frac{1}{3}S \gamma_{cd}\right).
\end{equation}
If the two regions on either side of the brane are identical, a ${\mathbb Z}_2$ symmetry is imposed which leads to $K^+_{cd}= -K^-_{cd}$ and $[K_{cd}]=2K^+_{cd}$. Considering that the brane energy momentum tensor contains only the brane tension $S_{cd}=-\sigma \gamma_{cd}$, the junction condition reduces to,
\begin{equation}\label{ext_curv}
	K^+_{cd}=-\frac{\kappa^2}{6}\sigma \gamma_{cd}.
\end{equation}
Using the value of spatial components of (\ref{extrcom}) in (\ref{ext_curv}), we get,
\begin{equation}\label{vsig}
\frac{\sqrt{f(a)+\dot{a}^2}}{a}=-\frac{\kappa^2}{6}\sigma.
\end{equation}
Substituting the value of $f(a)$ from equation (\ref{fr}) in the above equation, the first Friedmann equation can be written in the following form,
\begin{equation}\label{FRW_eqn}
H^2=\frac{\dot{a}^2}{a^2}=-\frac{k}{a^2}+\frac{m}{a^4}+\frac{2b}{3a^3}+\frac{\Lambda_4}{3},
\end{equation}
where, $H$ represents the Hubble factor and $\Lambda_4=\frac{\kappa^4\sigma^2}{12}-\frac{3}{l^2}$ is the cosmological constant on the brane. By tuning the brane tension and AdS curvature radius, the value of the brane cosmological constant can be set as positive, negative or zero accordingly. Here, the effective Friedmann equation involves a mass parameter term $m$ that gives a dark radiation contribution and a string density term $b$ that gives an effective matter-like contribution induced by the bulk geometry.

Differentiating the Friedmann equation with respect to $\tau$, we obtain the second Friedmann equation as,
\begin{equation}\label{RCeqn}
\frac{\ddot{a}}{a}=-\frac{m}{a^4}-\frac{b}{3a^3}+\frac{\Lambda_4}{3}.
\end{equation}
Solving these Friedmann equations we analyze the nature of cosmological evolution of the brane universe.  The nature of solution of equation (\ref{FRW_eqn}) depends on the mass of the black hole, nature of curvature of spacetime and cosmological constant on the brane. We study all the possible combinations of these parameters individually.
\subsection*{(I) $\Lambda_4 =0$ and $m \geq 0$:}
\begin{figure}[h]
\centering
\includegraphics[width=0.45\linewidth]{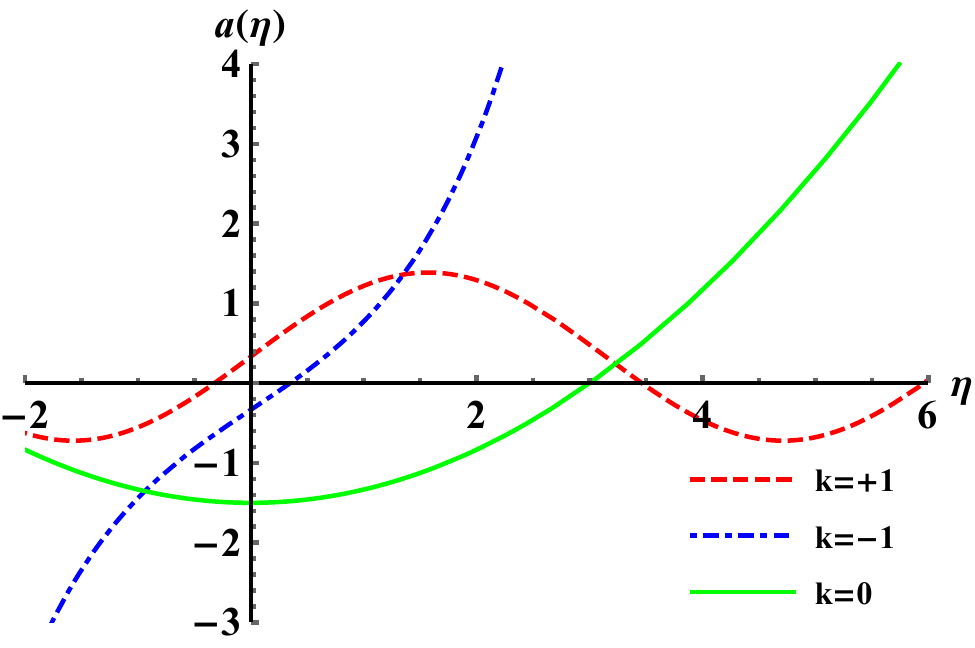}
\hspace{.2in}
\caption{Plot of $a$ vs $\eta$ for different values of $k$, $m=1$ and $b=1$. 
 }
\label{fig 2}
\end{figure}

We first analyze the case of a critical brane ($\Lambda_4=0$) by setting $l=3/\sigma$ and use the conformal time $\eta$, defined as $d\tau=a(\eta)d\eta $. The solutions of the Friedmann equation are found to be,
\begin{align}
a(\eta)=&\left(m+\frac{b^2}{9}\right)^{\frac{1}{2}}\sin\eta+\frac{b}{3}  \hspace{1cm}\text{for} \hspace{0.2 cm}k=+1,\\
=&\left(m-\frac{b^2}{9}\right)^{\frac{1}{2}}\sinh\eta-\frac{b}{3}\hspace{0.85cm}\text{for} \hspace{0.2 cm}k=-1,\\
=& \hspace{0.1cm}\frac{b}{6}\eta^2-\frac{3m}{2b}\hspace{3.25cm}\text{for} \hspace{0.2 cm}k=0.
\end{align}
The nature of the scale factor $a(\eta)$ as a function of the conformal time $\eta$ is plotted in figure (\ref{fig 2}). For a closed universe ($k=1$), the scale factor posses a non-zero value at $\eta=0$, but collapses back to singularity after a brief period of expansion.
For an open universe $k=-1$ and for flat universe $k=0$, the scale factor undergoes an exponential and a quadratic expansion respectively.  
\subsection*{(II) $\Lambda_4 \ne 0$ and $m \geq 0$:}
\begin{figure}[h]
	\centering
\subfigure[]{\includegraphics[width=0.475\linewidth]{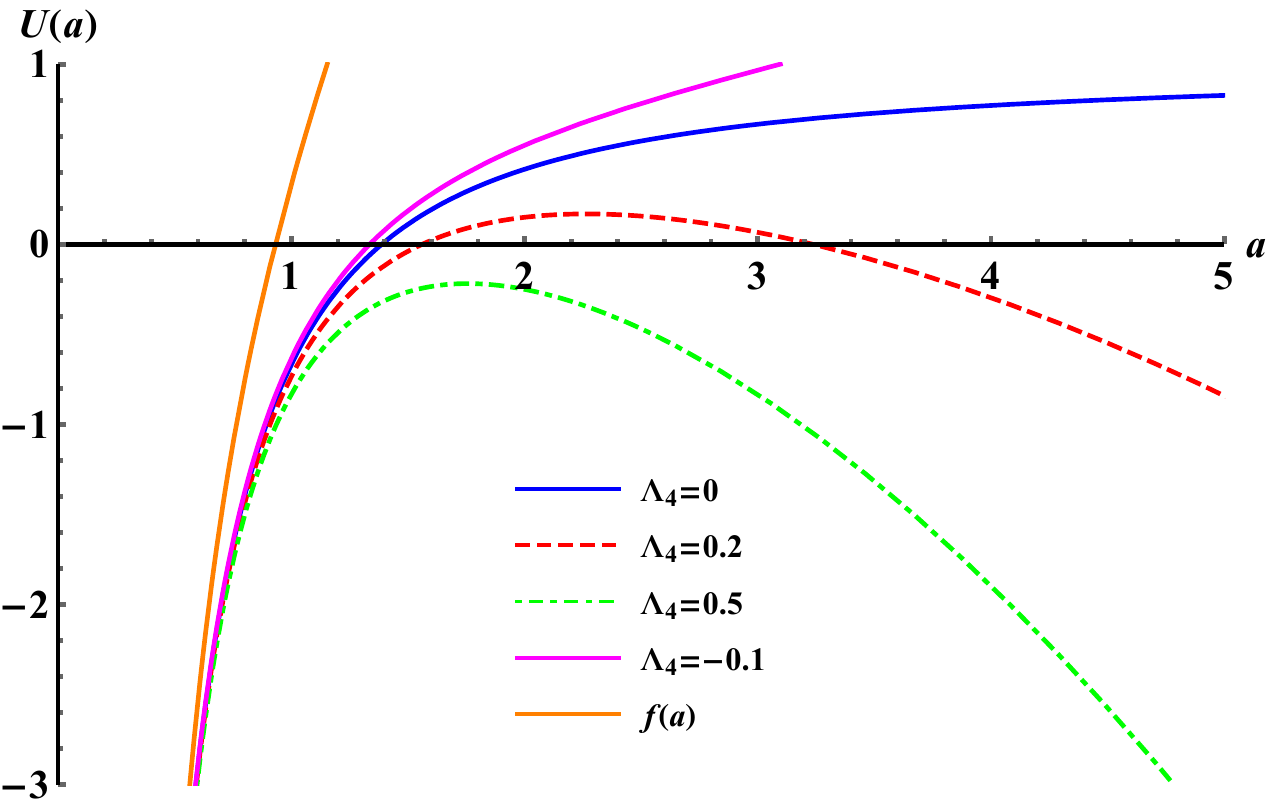}}
\hspace{.2in}
\subfigure[]{\includegraphics[width=0.475\linewidth]{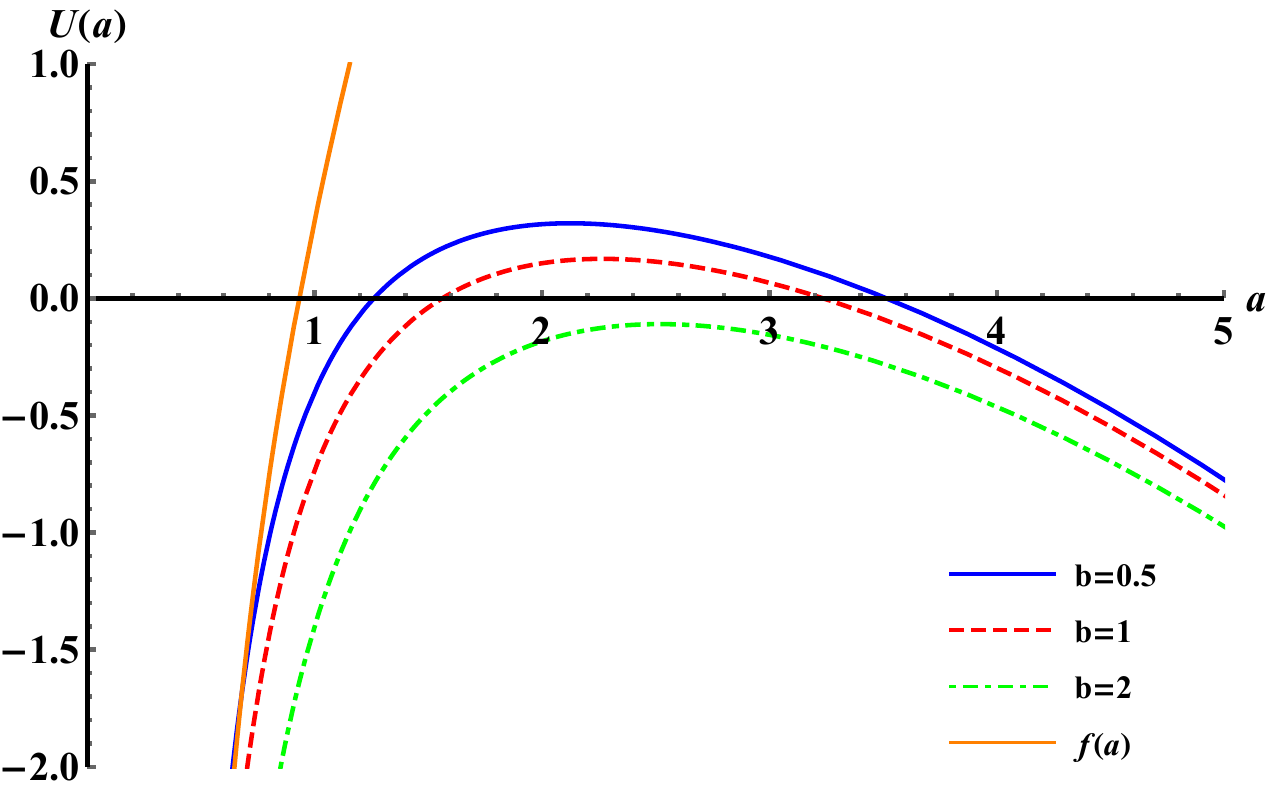}}
\caption{Plot of $U(a)$ with respect to $a$ with $k=+1$ for different values of (a) $\Lambda_4 (m=1, b=1)$ and (b) string density $(m=1,  \Lambda_4=0.2 )$. Here, $m=1,b=1,l=1$ for $f(a)$. }
\label{fig 3}
\end{figure}

An analytical solution of equation (\ref{FRW_eqn}) for a non-critical brane is difficult to obtain. However, we can perform a qualitative analysis of the nature of the cosmological evolution by treating the world volume of brane as a classical particle and the corresponding Hamilton's constraint equation is considered to be the evolution equation of the brane. The evolution equation can be constructed by rewriting the equation (\ref{vsig}) as,
\begin{equation}\label{aU}
\left( {\frac{da}{d\tau}}\right)^2+U(a)=0,
\end{equation}
where,
\begin{align}
\label{ua}
U(a)&=f(a)-\frac{\kappa^2\sigma^2}{36}a^2 = k-\frac{m}{a^2}-\frac{2b}{3a}-\Lambda_4 \frac{a^2}{3}.
\end{align}
The first term of the equation (\ref{aU}), is proportional to the kinetic
energy and the second term can be considered as an effective potential of the brane world
having total energy equal to zero. When $U(a)> 0$, the kinetic energy of the universe becomes negative and the universe can not exist in this region and for $U(a) = 0$, the rate of expansion/contraction of the universe $\dot a = 0$, representing the turning point of the movement of the brane world along $a$. Therefore, $U(a)\leq0$ represents the physical domain of the solution. We have analysed the dynamical nature of world volume by plotting $U(a)$ against $a$ in figure (\ref{fig 3}) for various values of $k$ and $\Lambda_4$ and expressed the equation (\ref{ua}) as the polynomial of $a$ in order to determine the number of solutions of $U(a)=0$. The polynomial is
\begin{equation}
\Lambda_4 a^4 -3k a^2 + 2 ba + 3m =0. 
\end{equation}
 For $k=+1$ and $\Lambda_4>0$, Descarte's sign rule indicates that there are maximum two real positive roots or no roots. In case of two roots, there are two values of $a$ for which the universe expansion or contraction rate becomes zero. The plots in figure (\ref{fig 3}) shows that, within a certain small range of $\Lambda_4$ and below a certain critical value of string density, $U(a)$ has one maximum within the two roots.  So there are two possible region for the universe to exist. One region has a minimum size equal to the larger root $a_b$ (say point of bounce of universe). From this size, the universe bounces back from a contracting phase to an expanding phase which leads to a nonsingular universe. This bounce is induced by the spatial curvature and occurs at large $a$ even for $m=b=0$. Also, the brane must bounce outside the horizon of the black hole in order to obtain a stable bounce as discussed in \cite{DeRisi:2007dn}. In figure (\ref{fig 3}) we have plotted $f(a)$ along with the plots of $U(a)$ which shows that the bounce occurs outside the horizon of the black hole. In the other possible region, the universe can have maximum size equal to the smaller root $a_c$. In this case, universe crunches from an expanding mode to a contracting mode and finally goes to a singular point which provides a singular universe. As $\Lambda_4$ and string density increases the two roots approach to each other and merge at the critical value of $\Lambda_4$ and string density. Above the critical values, $U(a)$ has no roots implying the universe either expands eternally without crunching or contracts to a singularity. This reflects the fact of transition of three black hole solutions to a single black hole solution in five dimensional bulk spacetime \cite{Dey:2017xty}. In the case of $k=+1$ and $\Lambda_4<0$, there is only one real positive root of $U(a)=0$ indicating that the universe quickly turn back from expanding mode to contracting mode and finally collapses to the singularity.

\begin{figure}[H]
	\centering
\subfigure[]{\includegraphics[width=0.475\linewidth]{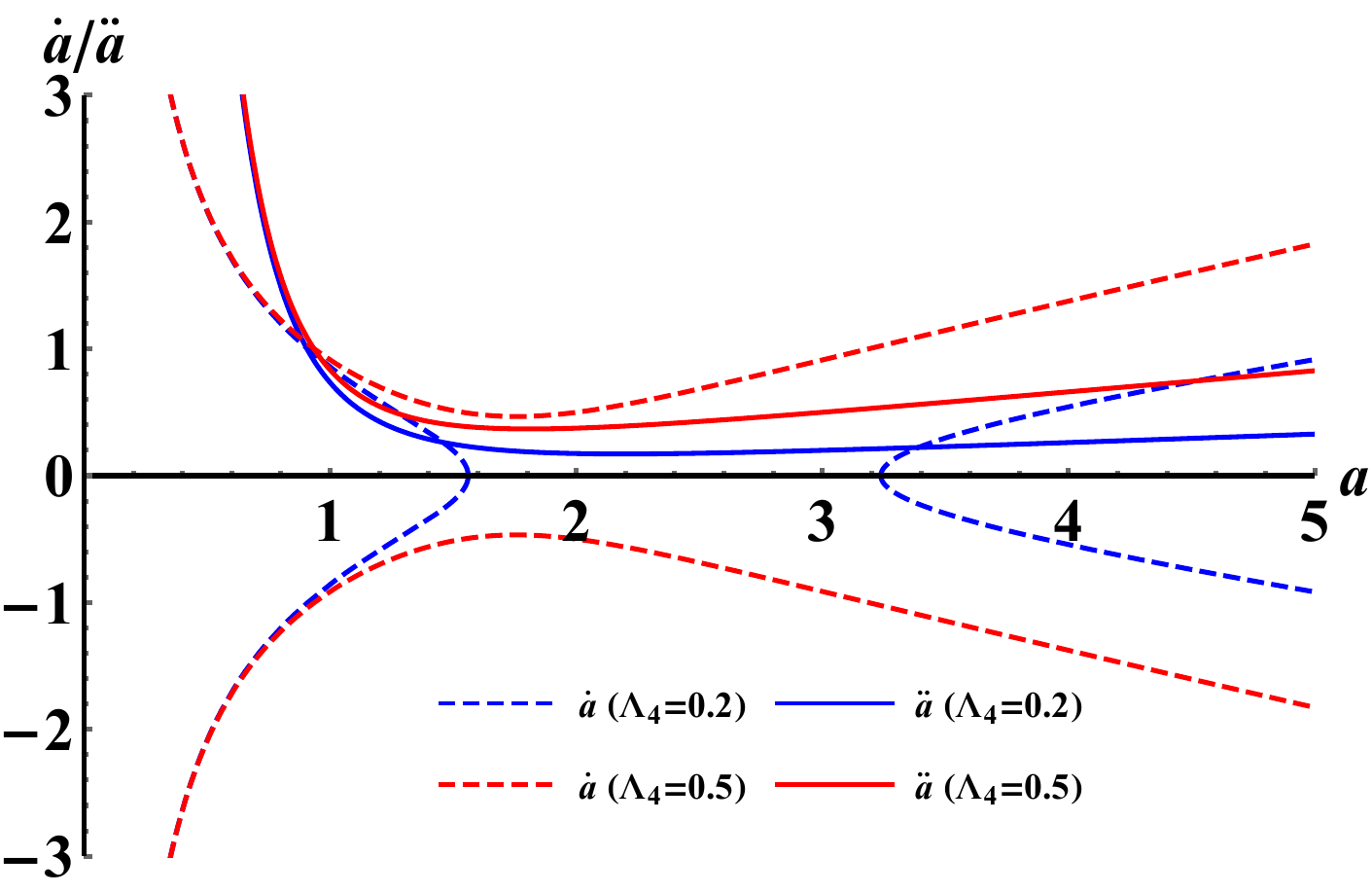}}
\hspace{.2in}
\subfigure[]{\includegraphics[width=0.475\linewidth]{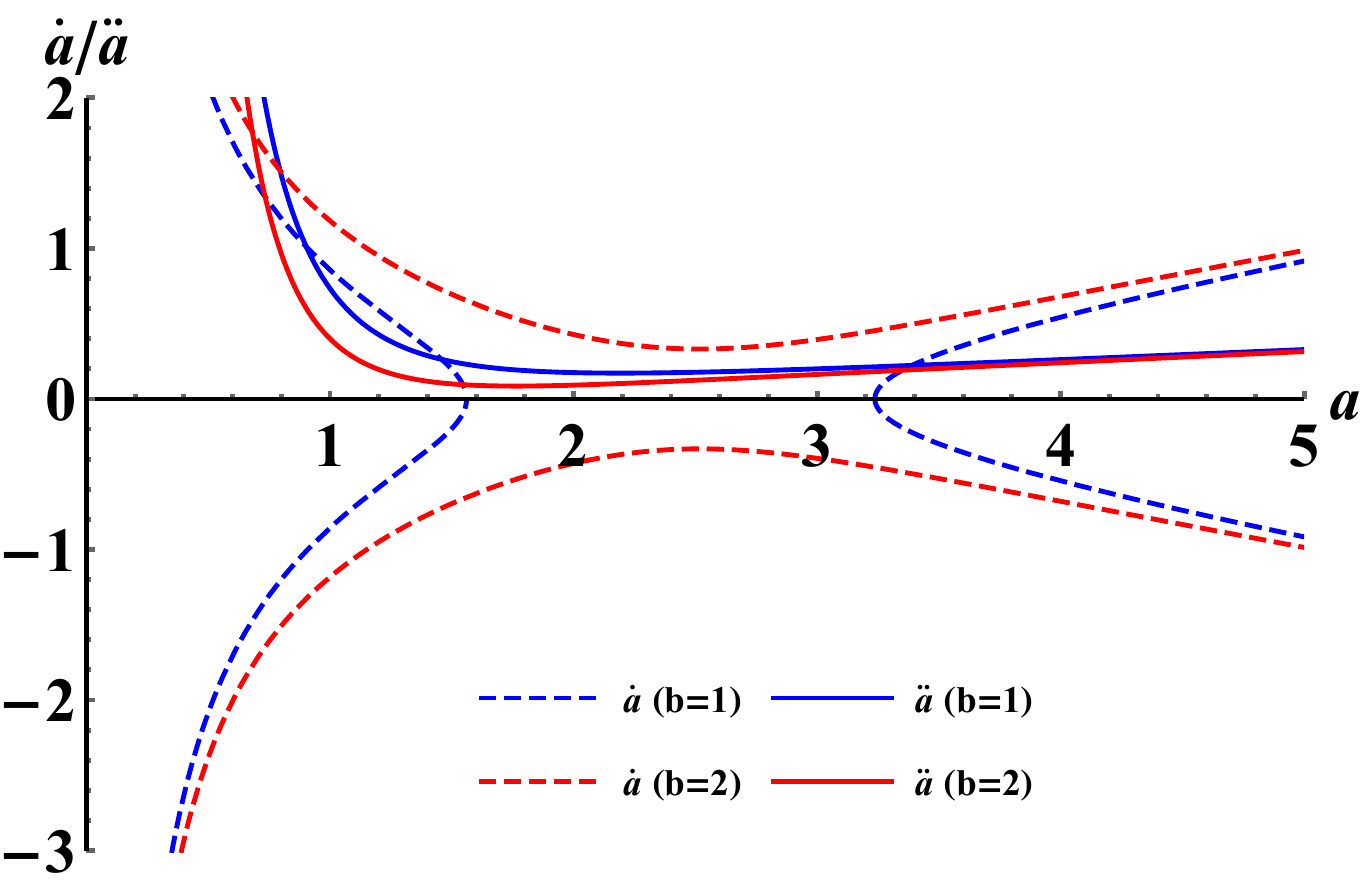}}
\caption{Variation of $\dot{a}$ and $\ddot{a}$ with respect to $a$ for various values of (a) $\Lambda_4 (k=+1,m=1,b=1)$ and (b) string density $(k=+1,m=1,\Lambda_4=0.2)$. }
\label{fig 4}
\end{figure}

The conditions for a bounce to occur at the bouncing point $a=a_b$ are $\dot{a}=0$ and $\ddot{a}>0$. Motivated by figure (\ref{fig 3}), we draw figure (\ref{fig 4}), where $\dot{a}$ and $\ddot{a}$ are plotted with respect to $a$ for both the cases of different values of $\Lambda_4$ and $b$. We find that the conditions for a bounce are satisfied at $a=a_b$.

\begin{figure}[H]
	\centering
\subfigure[]{\includegraphics[width=0.45\linewidth]{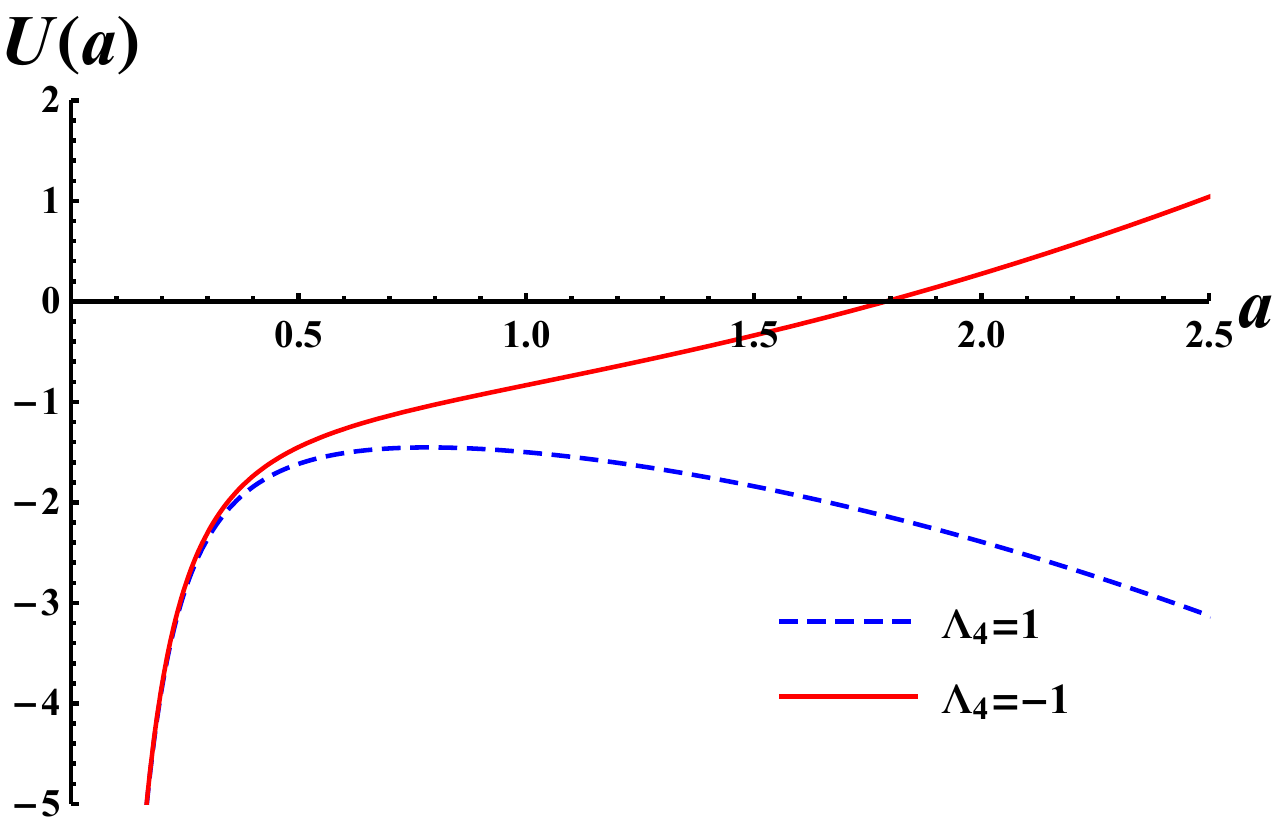}}
\hspace{.2in}
\subfigure[]{\includegraphics[width=0.45\linewidth]{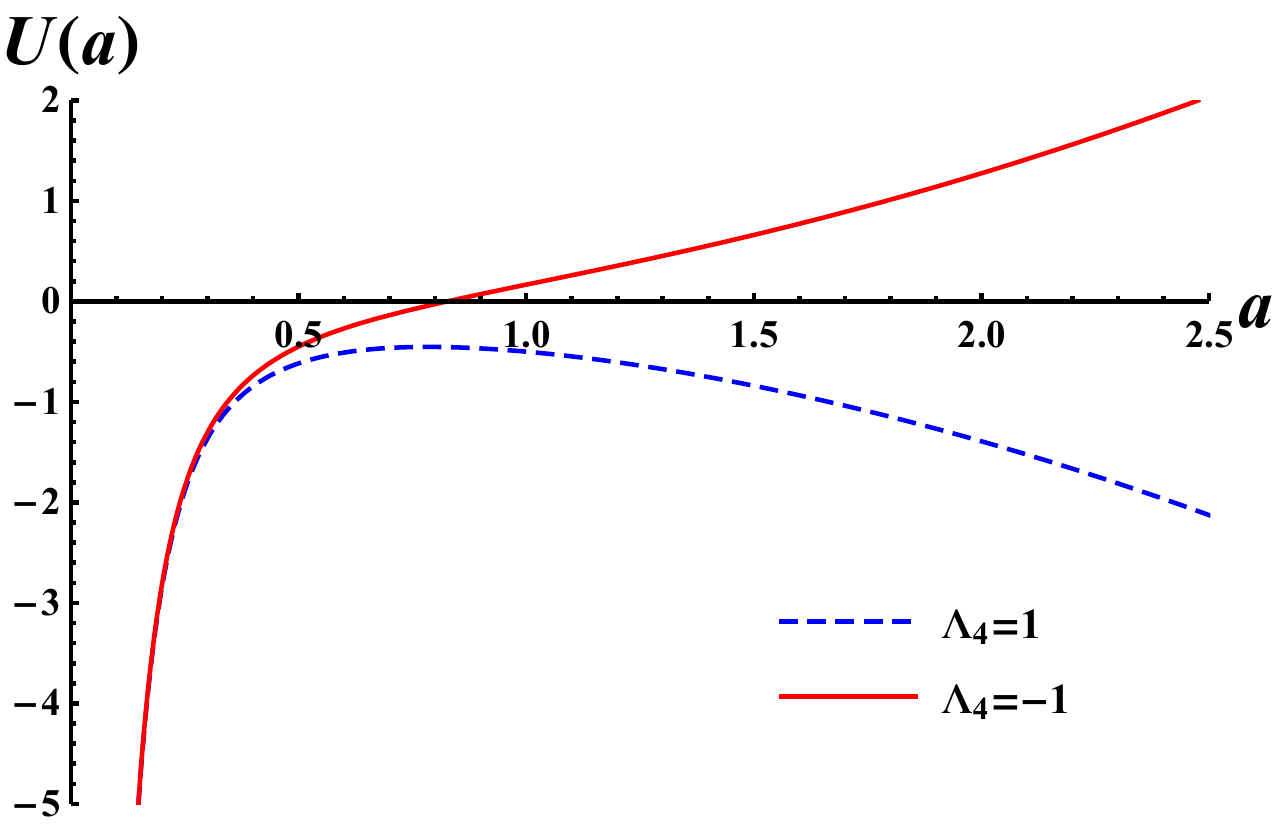}}
\caption{Plot of $U(a)$ with respect to $a$ for various values of $\Lambda_4$, $ m=0.1$, $b=0.1$ with (a) $ k=-1$ and (b)$ k=0$. }
	\label{fig 5}
\end{figure}

For $k=0$ or $-1$, there is no real positive root for $U(a)=0$ in the case of $\Lambda_4>0$ as shown in figure (\ref{fig 5}). Therefore, the universe does not have any bouncing or crunching nature. For $\Lambda_4<0$, there is only one real positive root of $U(a)=0$ showing that the universe quickly turns back from expanding mode to contracting mode and finally collapses to the singularity.

\subsection*{(III) $\Lambda_4 =0$ and $m<0$:}
\begin{figure}[H]
	\centering
\subfigure[]{\includegraphics[width=0.45\linewidth]{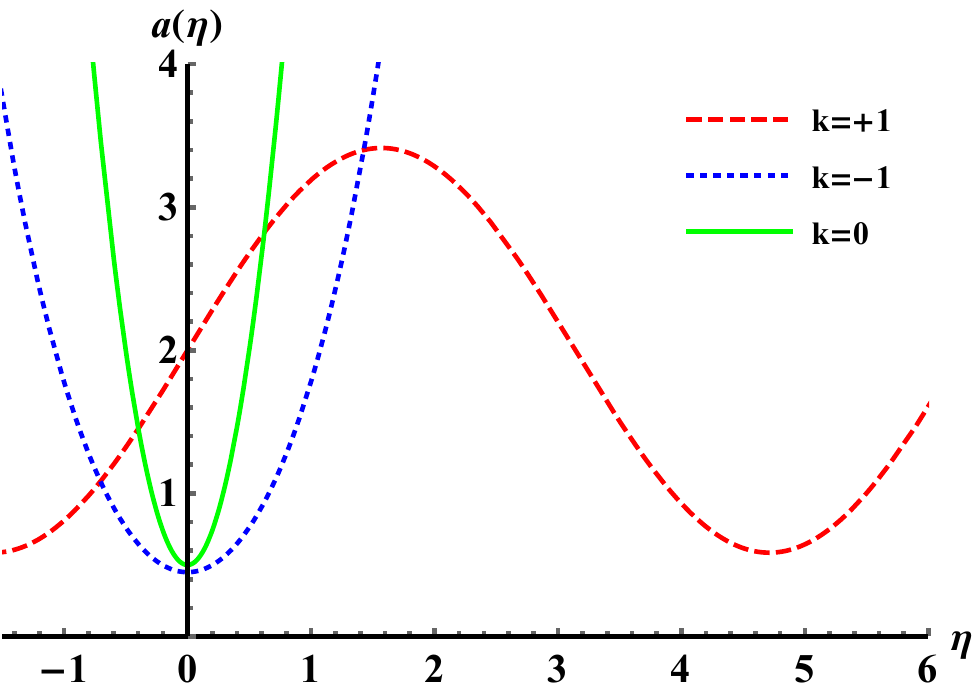}}
\hspace{.2in}\subfigure[]{\includegraphics[width=0.45\linewidth]{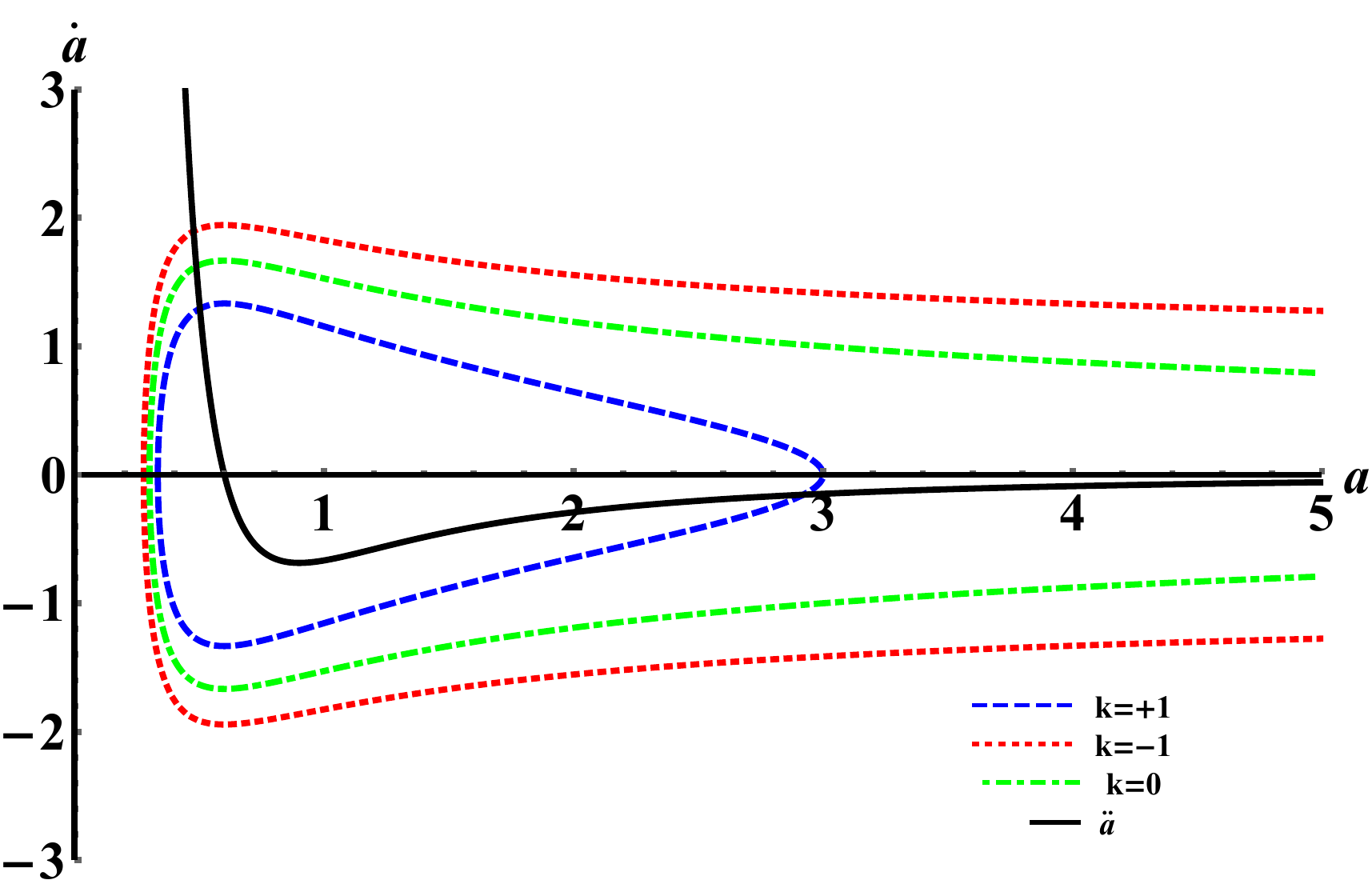}}
\caption{(a) Plot of $a$ vs $\eta$ for different values of $k$, $m=2$ and $b=6$. (b)Plot of $\dot a$  and $\ddot a$ vs $a$ for different values of $k$, $m=1$ and $b=5$. }
\label{fig 6}
\end{figure}

From equation (\ref{mass}), it is evident that the bulk can have a negative mass parameter at smaller values of the bulk radius on increasing values of the string density. This negative mass parameter induces a negative dark radiation contributing to a negative energy density. We now analyze the cosmological evolution of the bulk with a negative mass parameter. Friedmann equations (\ref{FRW_eqn}) and (\ref{RCeqn}) are modified as 
\begin{equation}\label{FRW_eqn1}
H^2=\frac{\dot{a}^2}{a^2}=-\frac{k}{a^2}-\frac{m}{a^4}+\frac{2b}{3a^3}+\frac{\Lambda_4}{3},
\end{equation}
and 
\begin{equation}\label{RCeqn1}
\frac{\ddot{a}}{a}=\frac{m}{a^4}-\frac{b}{3a^3}+\frac{\Lambda_4}{3}.
\end{equation}
In case of a critical brane, the solutions of the Friedmann equation (\ref{FRW_eqn}) are found to be,
 \begin{align}
 	a(\eta)=&\left(\frac{b^2}{9}-m\right)^{\frac{1}{2}}\sin\eta+\frac{b}{3}  \hspace{1cm}\text{for} \hspace{0.2 cm}k=+1,\\
 	=&\left(m+\frac{b^2}{9}\right)^{\frac{1}{2}}\cosh\eta-\frac{b}{3}\hspace{0.85cm}\text{for} \hspace{0.2 cm}k=-1,\\
 	=& \hspace{0.1cm}\frac{b}{6}\eta^2+\frac{3m}{2b}\hspace{3.25cm}\text{for} \hspace{0.2 cm}k=0.
 \end{align}
For a closed universe ($k=1$) and $m <\frac{b^2}{9}$, we find that the scale factor exhibits a nonsingular bouncing and cyclic nature with a minimal and maximal size given by,
\begin{equation}
	a_{max}=\left[\frac{b^2}{9}-m\right]^{\frac{1}{2}}+\frac{b}{3} \hspace{1cm} \text{and}\hspace{1cm} a_{min}=-\left[\frac{b^2}{9}-m\right]^{\frac{1}{2}}+\frac{b}{3}.
\end{equation}
For an open universe $(k=-1)$, the scale factor undergoes an nonsingular exponential  expansion  with a minimum of the scale factor $\left[(b^2/9)-m\right]^{\frac{1}{2}}+b/3$ while for a flat universe $(k=0)$, the scale factor undergoes an nonsingular quadratic  expansion  with a minimum of the scale factor $m /b$. The nature of the scale factor $a(\eta)$ with respect to conformal time $\eta$ is plotted in figure (\ref{fig 6}a). Similar nature is observed from the plot of $\dot a$ and $\ddot a$ with respect to size $a$ in figure (\ref{fig 6}b). 

\subsection*{(IV) $\Lambda_4 \ne 0$ and $m<0$:}
\begin{figure}[H]
\centering
\subfigure[]{\includegraphics[width=.45\linewidth]{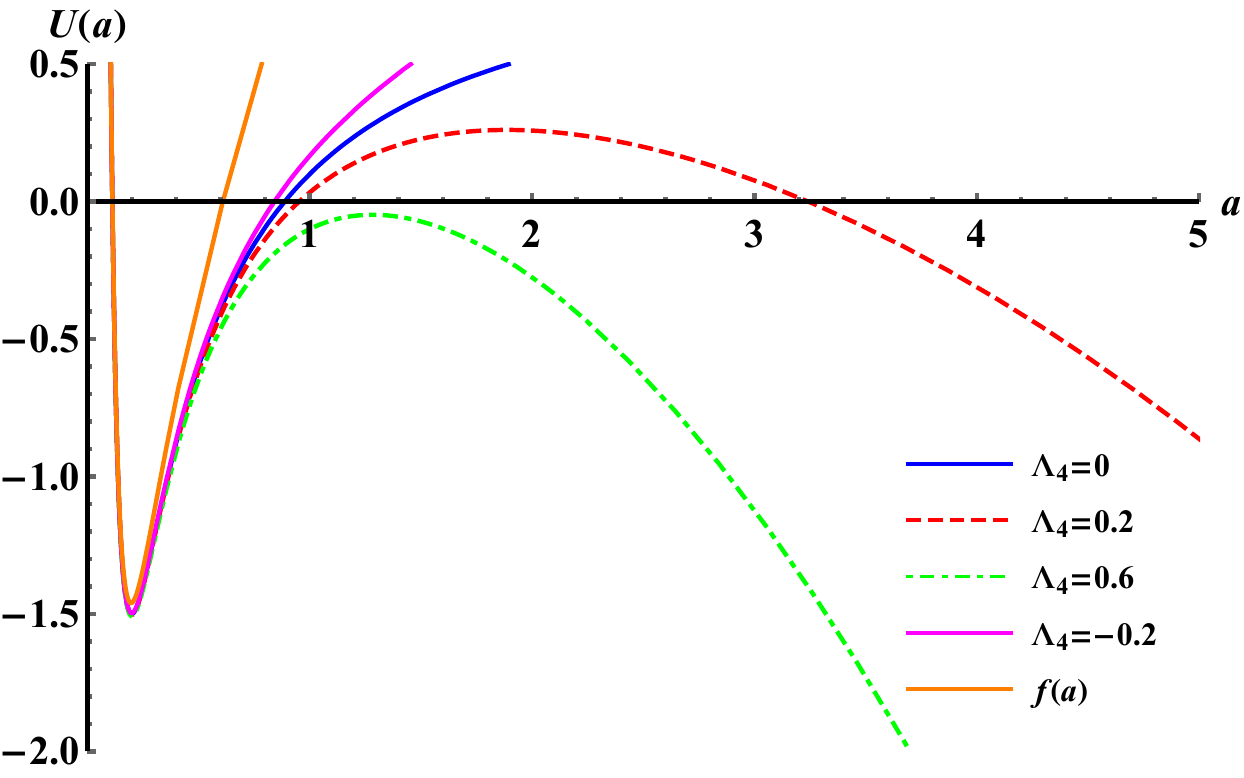}}
\hspace{.2in}
\subfigure[]{\includegraphics[width=.45\linewidth]{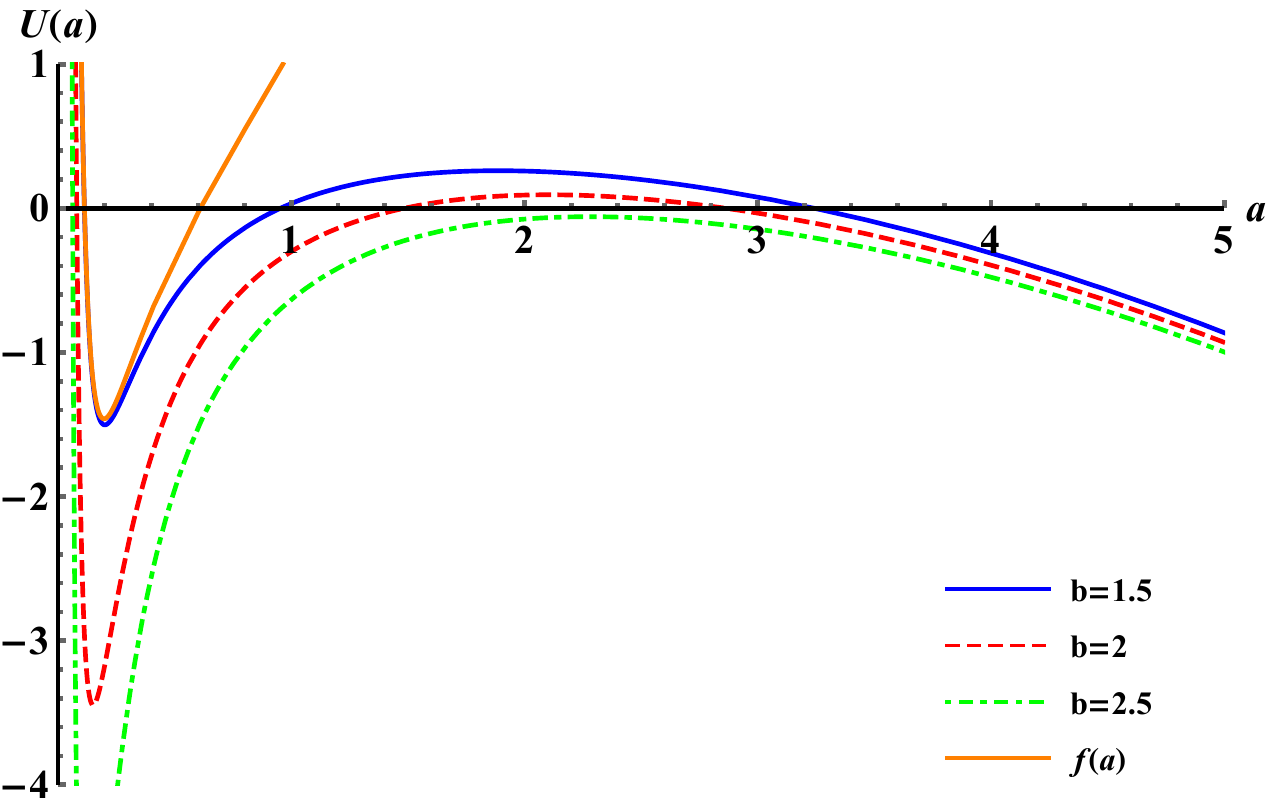}}
\caption{Plot of $U(a)$ as a function of $a$ with $k=+1$ for different values of (a) $\Lambda_4 (m=0.1, b=1.5)$ and (b) string density $(m=0.1, \Lambda_4=0.2)$. Here, $ k=+1, m=0.1, l=1, b=1.5$ for $f(a)$. }
\label{fig 7}
\end{figure}

In order to understand the cosmological evolution of the non-critical brane, we follow the qualitative analysis method as discussed in the previous case using equation (\ref{aU}). In the case of a negative mass parameter, the polynomial equation of $a$ for $U(a)=k+\frac{m}{a^2}-\frac{2b}{3a}-\Lambda_4 \frac{a^2}{3}=0$ can be arranged as,
\begin{equation}
\Lambda_4 a^4 -3k a^2 + 2 ba - 3m =0. 
\end{equation}
For $k=+1$ and $\Lambda_4>0$, Descarte's sign rule demands that there are maximum three real positive roots or minimum one real positive root. As observed from the plots in figure (\ref{fig 7}), there are two possible regions for the universe to exist within a certain range of $\Lambda_4$ and below a certain critical value of string density. We obtain a nonsingular universe driven by the spatial curvature corresponding to the largest root of $U(a)=0$ for $k=+1$.  The other two roots of $U(a)=0$ correspond to a bouncing and cyclic universe which crunches from expanding mode to contracting mode at the higher value of $a$ of the turning point and bounces back from contracting mode to expanding mode at the lower value of $a$ of the turning point. This bounce is induced by the negative energy density of the negative dark radiation term.
Above the critical value of $\Lambda_4$ and string density, only one region of universe exists where the universe begins from a nonsingular value of scale factor and expands continuously. Similar to the previous case, we have plotted $f(a)$ in figure (\ref{fig 7}) where it is shown that the black hole now has two horizons for a negative mass parameter. The bounce due to the dark radiation is inside the Cauchy horizon of the black hole and the bounce due to the spatial curvature is outside the black hole horizon. As discussed in \cite{Hovdebo:2003ug}, the bounce that occurs inside the Cauchy horizon is not feasible as it is unstable under perturbations. 
For $\Lambda_4<0$ with $k=+1$, there are two maximum real  positive roots of $U(a)=0$ specifying that the universe has a bouncing and cyclic nature.
\begin{figure}[H]
\centering
\subfigure[]{\includegraphics[width=.45\linewidth]{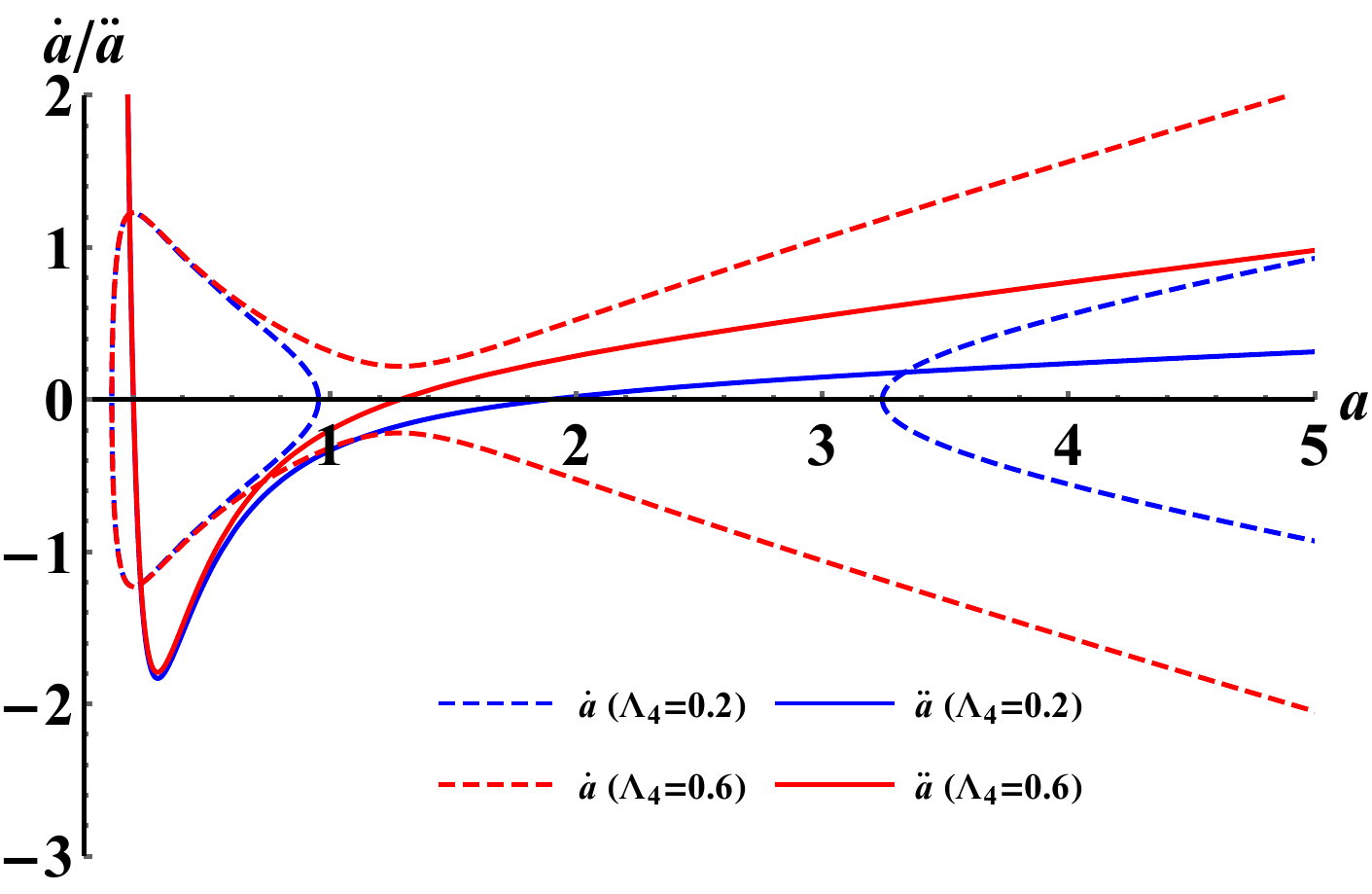}}
\hspace{.1in}
\subfigure[]{\includegraphics[width=.45\linewidth]{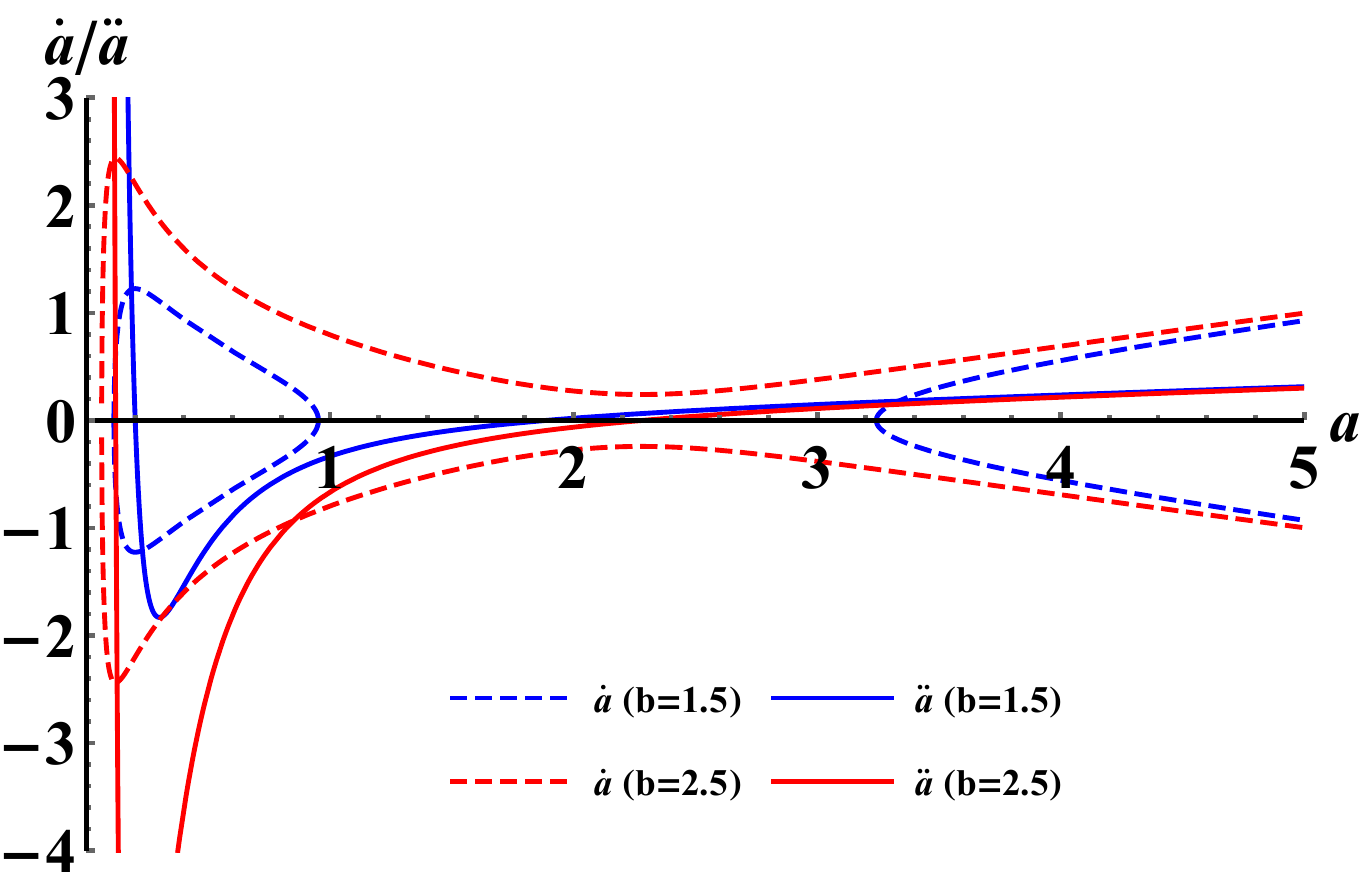}}
\caption{Plot of $\dot{a}$ and $\ddot{a}$ with respect to $a$ for various values of (a) $\Lambda_4 (m=0.1, b=1.5)$ and (b) string density $(m=0.1, \Lambda_4=0.2)$.}
\label{fig 8}
\end{figure}

The plot of $\dot{a}$ and $\ddot{a}$ with respect to the scale factor in figure (\ref{fig 8}) verifies the non-singular nature for various values of $\Lambda_4$ and string density.

\begin{figure}[H]
\centering
\subfigure[]{\includegraphics[width=.45\linewidth]{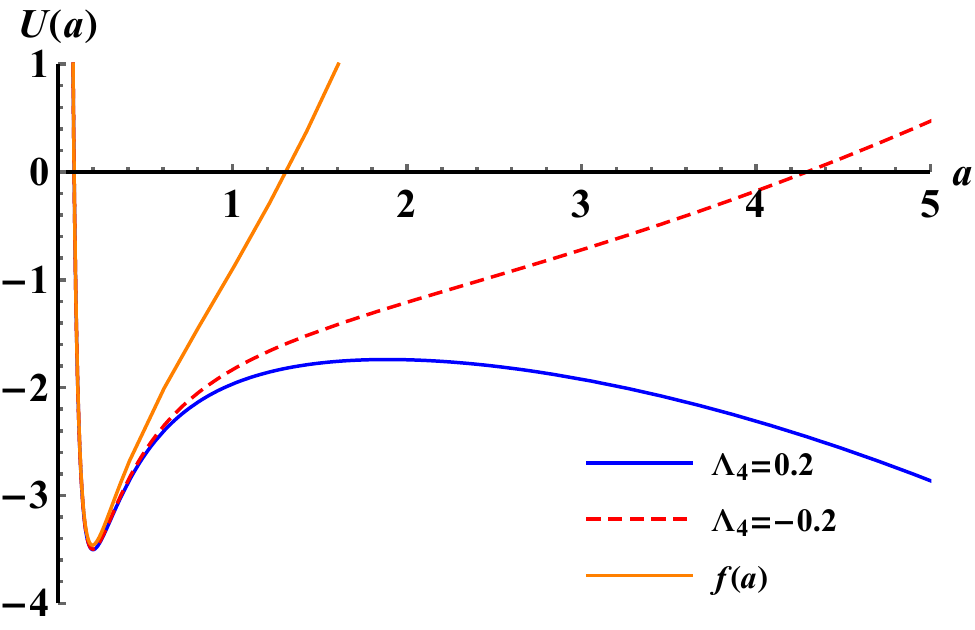}}
\hspace{.1in}
\subfigure[]{\includegraphics[width=.45\linewidth]{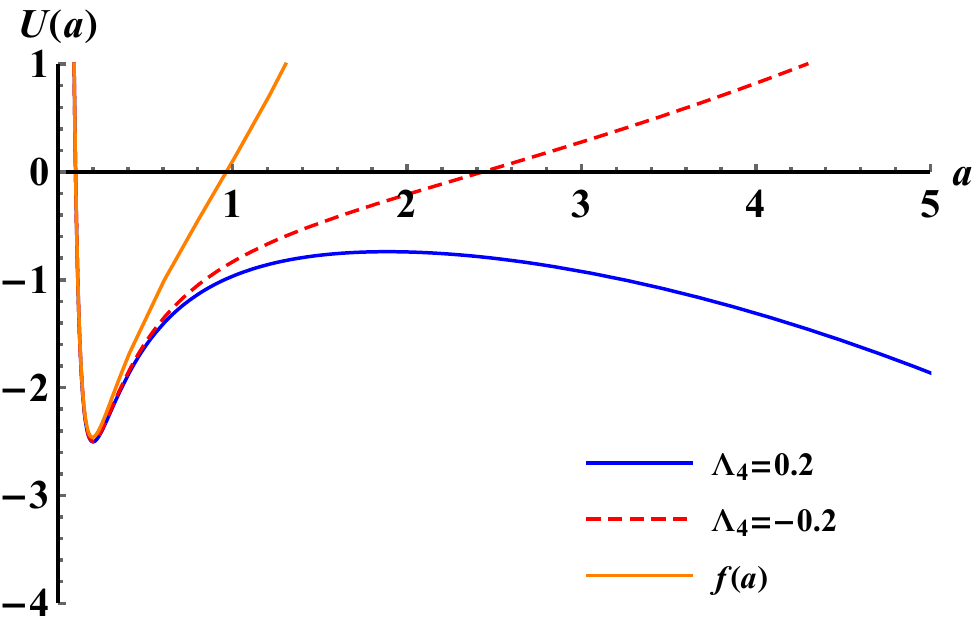}}
\caption{Plot of $U(a)$ with respect to $a$ for various values of $\Lambda_4$, $m=0.1$ and $b=1.5$ with  (a) $k=-1$ and (b) $k=0$.}
\label{fig 9}
\end{figure}

For $k=0$ and $-1$, in the case of a positive $\Lambda_4$, there is only one real positive root for $U(a)=0$. Thus, the universe gradually expands from a non-zero minimum size of the universe. In the case of a negative $\Lambda_4$, there are two maximum real positive roots of $U(a)=0$ suggesting that the universe has a bouncing and cyclic nature. The plots of $U(a)$ against $a$ for different values of $\Lambda_4$ with $k=0$ and $-1$ are shown in figure (\ref{fig 9}). However, since the bounce for both the cases of $k=0,-1$ occurs inside the horizons of the black hole, the bounce obtained is not feasible.

Thus, the bounce obtained in the braneworld scenario due to the spatial curvature is feasible as it occurs outside the horizon of the black hole. However, the bounce obtained  due to the negative dark radiation term is not feasible as it occurs inside the Cauchy horizon of the black hole. As a consequence, we now explore the possibility of stable bouncing scenario due to the negative dark radiation term in the dark bubble
(shellworld) model.

\subsection{The shellworld scenario}
 The shellworld lacks the $Z_2$ symmetry usually present in the RS braneworlds, such that one obtains an inside-outside configuration rather than the inside-inside configuration of the braneworld. The presence of mass and cloud of strings in the bulk results in the metric of the inside and outside regions to be,
\begin{equation}
ds^2_{\pm}=-f_{\pm}(r)dt^2+\frac{1}{f_{\pm}(r)}dr^2+r^2\tilde{\gamma}_{ij}dx^idx^j,
\end{equation}
where,	$+$ and $-$ signs represent the outside and inside regions respectively. The function $f_{\pm}(r)$ is given by the same metric function as in equation (\ref{fr}) with $k=1$ but for different mass parameters $m_{\pm}$, string densities $b_{\pm}$ and AdS curvature radii $l_{\pm}$ for the two regions. Here, $l_+$ is greater than $l_-$. Following the same approach as the braneworld scenario to calculate the extrinsic curvature, we find the spatial components of the extrinsic curvature on both sides of the shellworld as,
\begin{equation}	
K^{\pm}_{ij}=\frac{\sqrt{f_{\pm}(a)+\dot{a}^2}}{a}\gamma_{ij}.
\end{equation}
The junction condition (\ref{junction}) now reduces to,
\begin{equation}\label{ext_curv_shell}		
\frac{\sqrt{f_{-}(a)+\dot{a}^2}}{a}-\frac{\sqrt{f_{+}(a)+\dot{a}^2}}{a}	= \frac{\kappa^2}{3}\sigma.				
\end{equation}
The Friedmann equations for the shellworld model can be derived from the above junction condition. We now discuss the nature of evolution individually on the basis of sign of mass parameter.

\subsection*{(I) $m_{\pm} \geq 0$:}

The first Friedmann equation on the shellworld can be derived from the above equation as,
\begin{equation}\label{FRW_eqn_shell}	
\frac{\dot{a}^2}{a^2}=-\frac{1}{a^2}+\frac{1}{a^4}\left( \frac{ m_+l_+-m_-l_-}{l_+-l_-}\right)+\frac{2}{3a^3}\left( \frac{b_+l_+ - b_-l_-}{l_+-l_-}\right) +\frac{8\pi G_4}{3}\Lambda_4,
\end{equation}
where, $G_4=\frac{2}{l_+ - l_-} G_5$ is the four dimensional Newton's constant and $\Lambda_4\equiv\sigma_{crit}-\sigma$ is the cosmological constant on the shellworld. The brane behaves as a spatially flat Minkowski space with a critical tension when the five-dimensional energy scales exceed the four-dimensional ones \cite{Banerjee:2019fzz};
\begin{equation}
\sigma_{crit}=\frac{3}{8 \pi G_5}\left( \frac{1}{l_-}-\frac{1}{l_+}\right).
\end{equation} 
The second Friedmann equation on the shellworld can be derived by differentiating (\ref{FRW_eqn_shell}) with respect to $\tau$ and is given as,
\begin{equation}
\frac{\ddot{a}}{a}=-\frac{1}{a^4}\left( \frac{ m_+l_+-m_-l_-}{l_+-l_-}\right)-\frac{1}{3a^3}\left( \frac{b_+l_+ - b_-l_-}{l_+-l_-}\right) +\frac{8\pi G_4}{3}\Lambda_4.
\end{equation}
Using conformal time $\eta$, equation (\ref{FRW_eqn_shell}) can be solved exactly for $\Lambda_4=0$ to obtain the scale factor as,
\begin{equation}\label{scale_factor}
a(\eta)=\sqrt{\left( \frac{ m_+l_+-m_-l_-}{l_+-l_-}\right) + 
\frac{1}{9}\left( \frac{b_+l_+ - b_-l_-}{l_+-l_-}\right)^2} \sin \eta +\frac{1}{3}\left( \frac{b_+l_+ - b_-l_-}{l_+-l_-}\right).
\end{equation}
The solution is valid only when, $\left( \frac{ m_+l_+-m_-l_-}{l_+-l_-}\right) + \frac{1}{9}\left( \frac{b_+l_+ - b_-l_-}{l_+-l_-}\right)^2 \geq 0$. The maximum and minimum of the scale factor are given by,
\begin{align}	
a_{max}&=\sqrt{\left( \frac{ m_+l_+-m_-l_-}{l_+-l_-}\right) + \frac{1}{9}\left( \frac{b_+l_+ - b_-l_-}{l_+-l_-}\right)^2} +\frac{1}{3}\left( \frac{b_+l_+ - b_-l_-}{l_+-l_-}\right) \hspace{0.2cm}\text{and}\nonumber\\
a_{min}&=-\sqrt{\left( \frac{ m_+l_+-m_-l_-}{l_+-l_-}\right) + \frac{1}{9}\left( \frac{b_+l_+ - b_-l_-}{l_+-l_-}\right)^2} +\frac{1}{3}\left( \frac{b_+l_+ - b_-l_-}{l_+-l_-}\right).
\end{align}
A nonsingular bouncing and cyclic nature of the shellworld universe can be obtained under the conditions $  m_+l_+<m_-l_-$ and $ b_+l_+ > b_-l_- $. These particular conditions lead to a negative energy density contribution from the dark radiation term $(\sim a^{-4})$ which induces the bounce. The two conditions to get a singular and a nonsingular universe are plotted in figure (\ref{fig 11}a). The plot of $\dot{a}$ and $\ddot{a}$ in figure (\ref{fig 11}b) with respect to the size of the shellworld universe verifies the existence of nonsingular bouncing and cyclic nature of the scale factor.
\begin{figure}[h]
\centering
\subfigure{\includegraphics[width=0.45\linewidth]{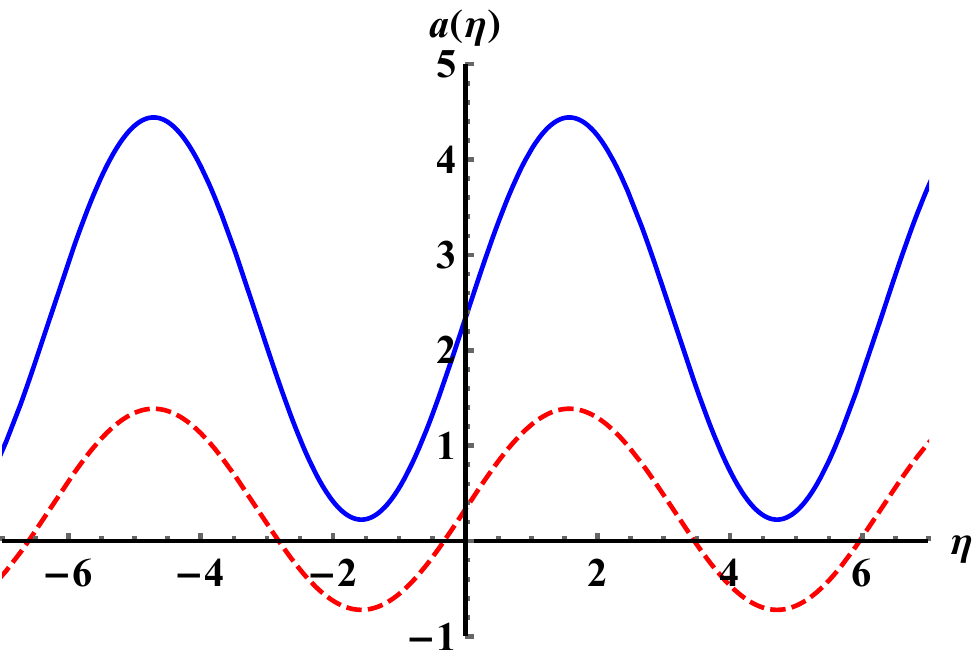}}	
\hspace{0.2cm}
\subfigure{\includegraphics[width=0.45\linewidth]{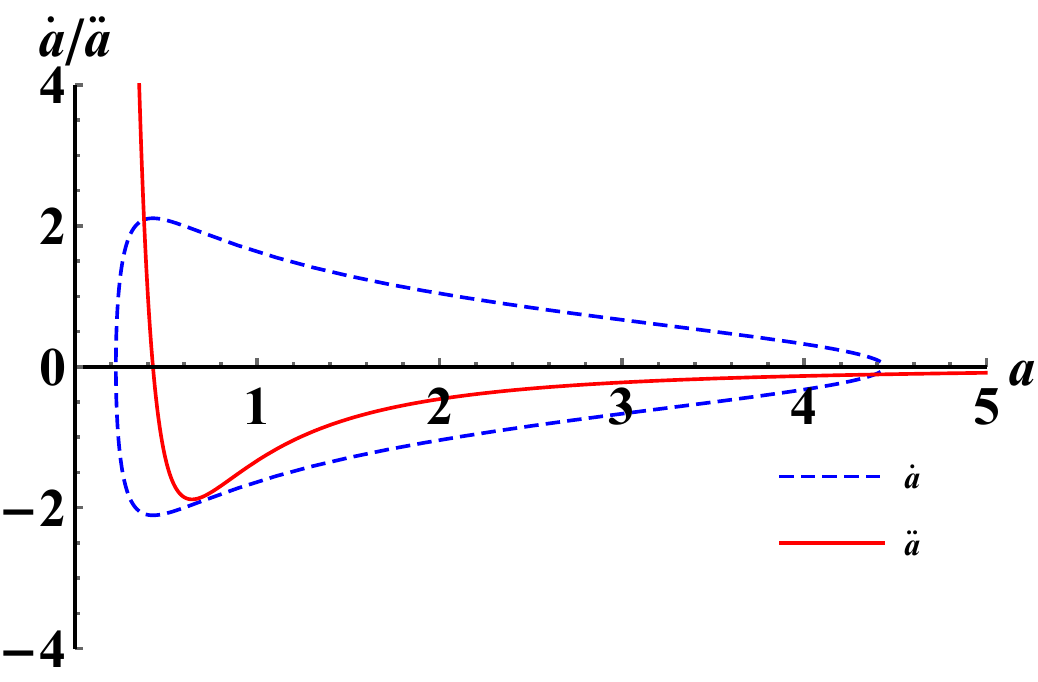}}
\caption{(a) Plot of $a$ vs $\eta$ showing singular and nonsingular nature of the shellworld universe. The blue curve represents the nonsingular universe with parameters $m_+=1$, $m_-=2$, $b_+=3$, $b_-=1$, $l_+=1.5$ and $l_-=1$. The red dashed curve represents the singular universe with parameters $m_+=1$, $m_-=1$, $b_+=1$, $b_-=1$, $l_+=1.5$ and $l_-=1$. (b) Plot of $\dot a$ and $\ddot a$ with respect to $a$ for the nonsingular universe with parameters $m_+=1$, $m_-=2$, $b_+=3$, $b_-=1$, $l_+=1.5$ and $l_-=1$. }
\label{fig 10}
\end{figure}

To study the nature of cosmological evolution of the shellworld universe in the presence of a non-zero brane cosmological constant, again the qualitative analysis method will be followed as discussed in the previous scenario. The effective potential of the shellworld universe is given by,
\begin{equation}\label{ua47}		
U(a)=1-\frac{1}{a^2}\left( \frac{ m_+l_+-m_-l_-}{l_+-l_-}\right)-\frac{2}{3a}\left( \frac{b_+l_+ - b_-l_-}{l_+-l_-}\right)-\frac{8\pi G_4}{3}\Lambda_4 a^2.
\end{equation}

We have plotted $U(a)$ against $a$ for the two conditions $( m_+l_+ \geq m_-l_-, b_+l_+ \geq b_-l_- $) and $( m_+l_+ < m_-l_-, b_+l_+ > b_-l_- )$ in figure (\ref{fig 11}) for different values of $\Lambda_4$ and string density. For the first condition, we find that a nonsingular nature of the shellworld universe is obtained for a certain range of $\Lambda_4$ and below a certain critical value of string density which is induced by the spatial curvature term and is outside the horizon of the black hole. For the second condition, we find two regions of the shellworld universe  where we can observe a nonsingular boucing nature. One is the bounce induced by the spatial curvature term which occurs at higher values of the scale factor within a certain range of $\Lambda_4$ and below a critical value of $b$ outside the horizon of the black hole, while the other is the bounce induced by the negative energy density of the dark radiation term which occurs inside the horizon of the black hole. 
\begin{figure}[H]
\centering

\subfigure[$m_+=m_-=1$, $b_+=b_-=1$, $l_+=1.5$ and $l_-=1$]{\includegraphics[width=.45\linewidth]{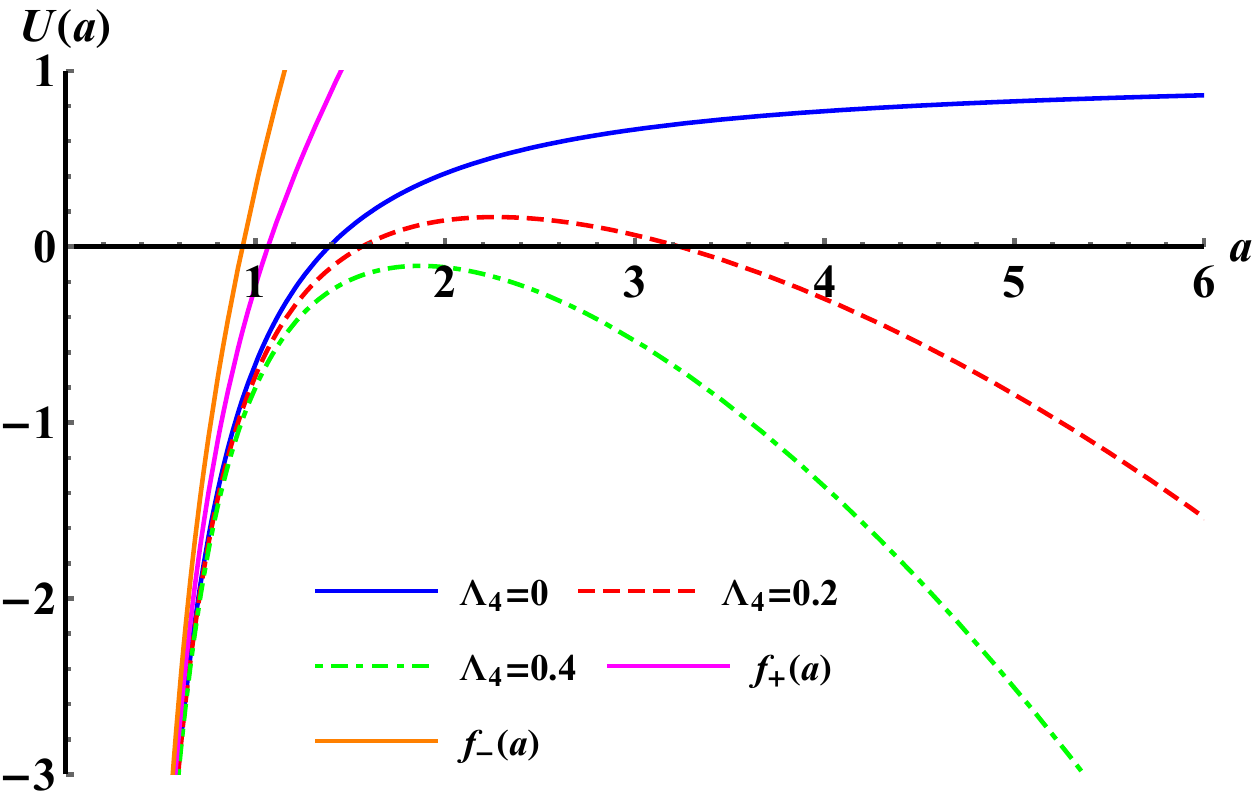}}
\hfill
\subfigure[$m_+=m_-=1$, $\Lambda_4=0.2$, $l_+=1.5$ and $l_-=1$]{\includegraphics[width=.45\linewidth]{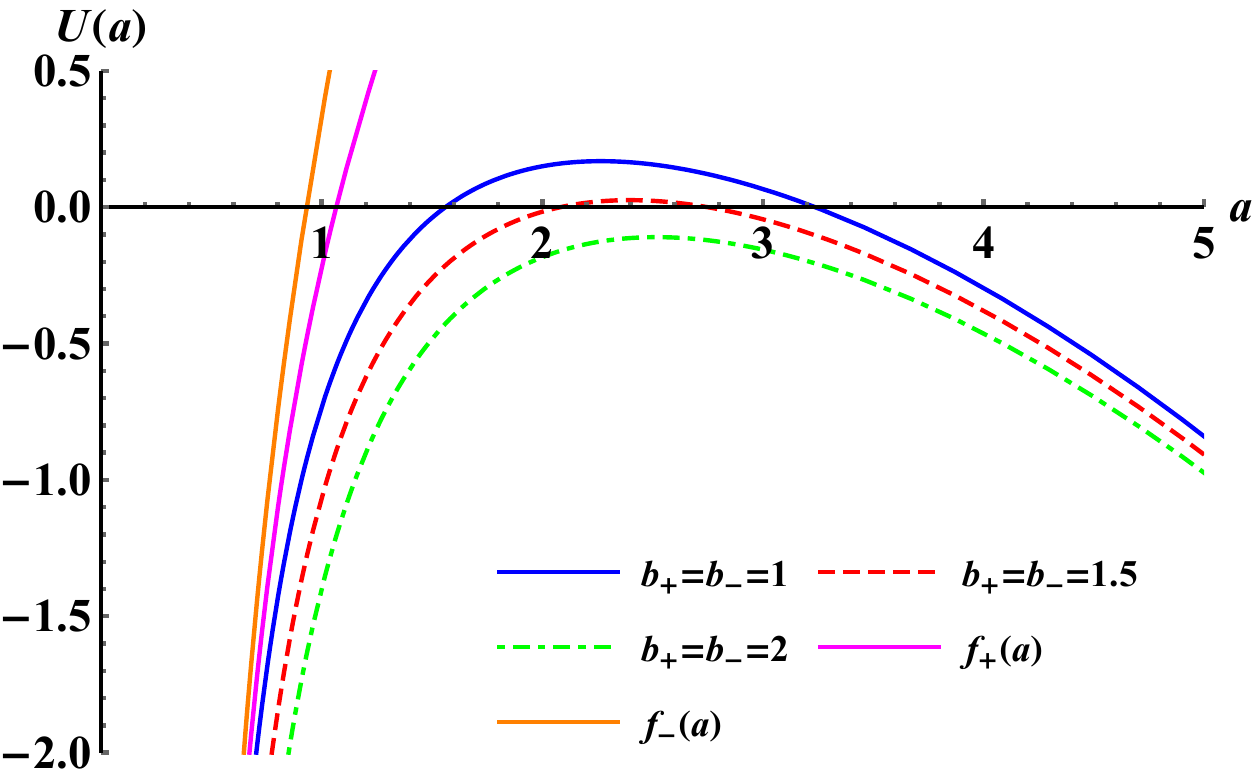}} \\[0.5cm]

\subfigure[ $m_+=0.5$, $m_-=1$, $b_+=b_-=3$, $l_+=1.5$ and $l_-=1$]{\includegraphics[width=.45\linewidth]{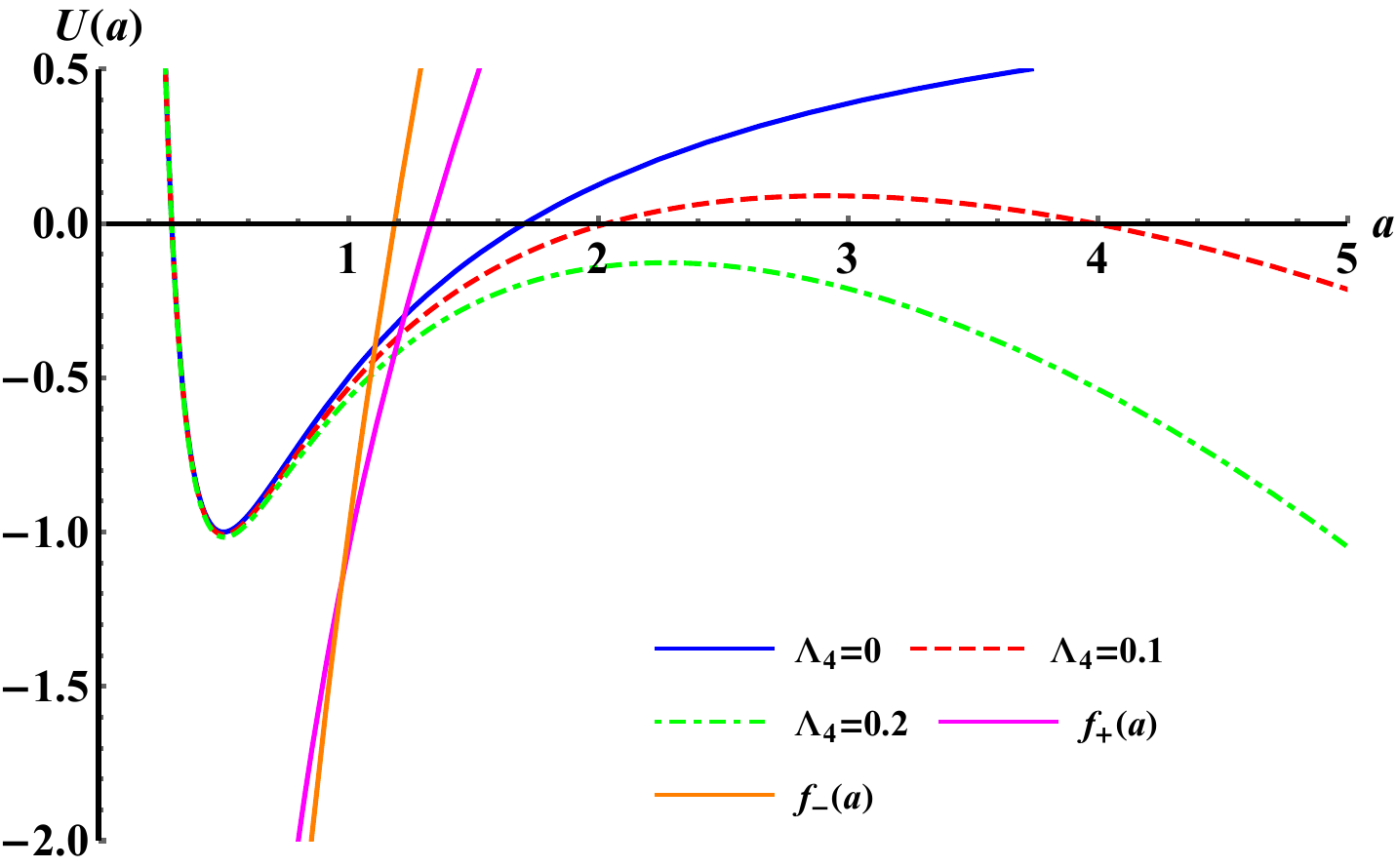}}
\hfill
\subfigure[ $m_+=0.5$, $m_-=1$, $\Lambda_4=0.1$, $l_+=1.5$ and $l_-=1$]{\includegraphics[width=.45\linewidth]{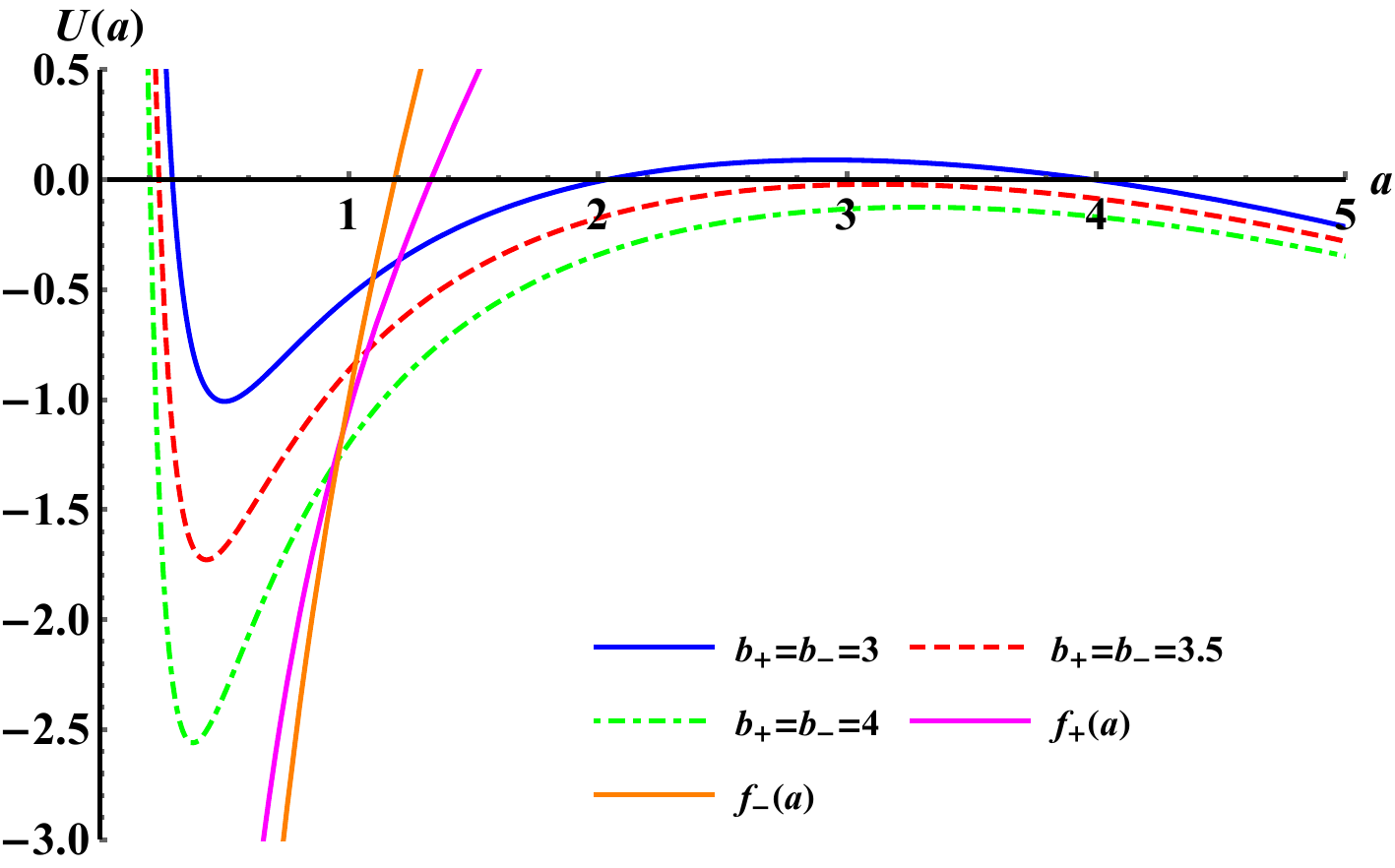}}

\caption{Plot of $U(a)$ and $f(a)$ against $a$ for different values of $\Lambda_4$ and string density. (a) and (b) correspond to the condition $m_+l_+>m_-l_-$ and (c) and (d) correspond to the condition $m_+l_+<m_-l_-$. }
\label{fig 11}
\end{figure}
The plot of $\dot{a}$ and $\ddot{a}$ against $a$ in figure (\ref{fig 12}) for the two conditions verifies the nonsingular bouncing nature of the shellworld universe as discussed above.

\begin{figure}[h]
\centering

\subfigure[$m_+=m_-=1$, $b_+=b_-=1$, $l_+=1.5$ and $l_-=1$]{\includegraphics[width=.45\linewidth]{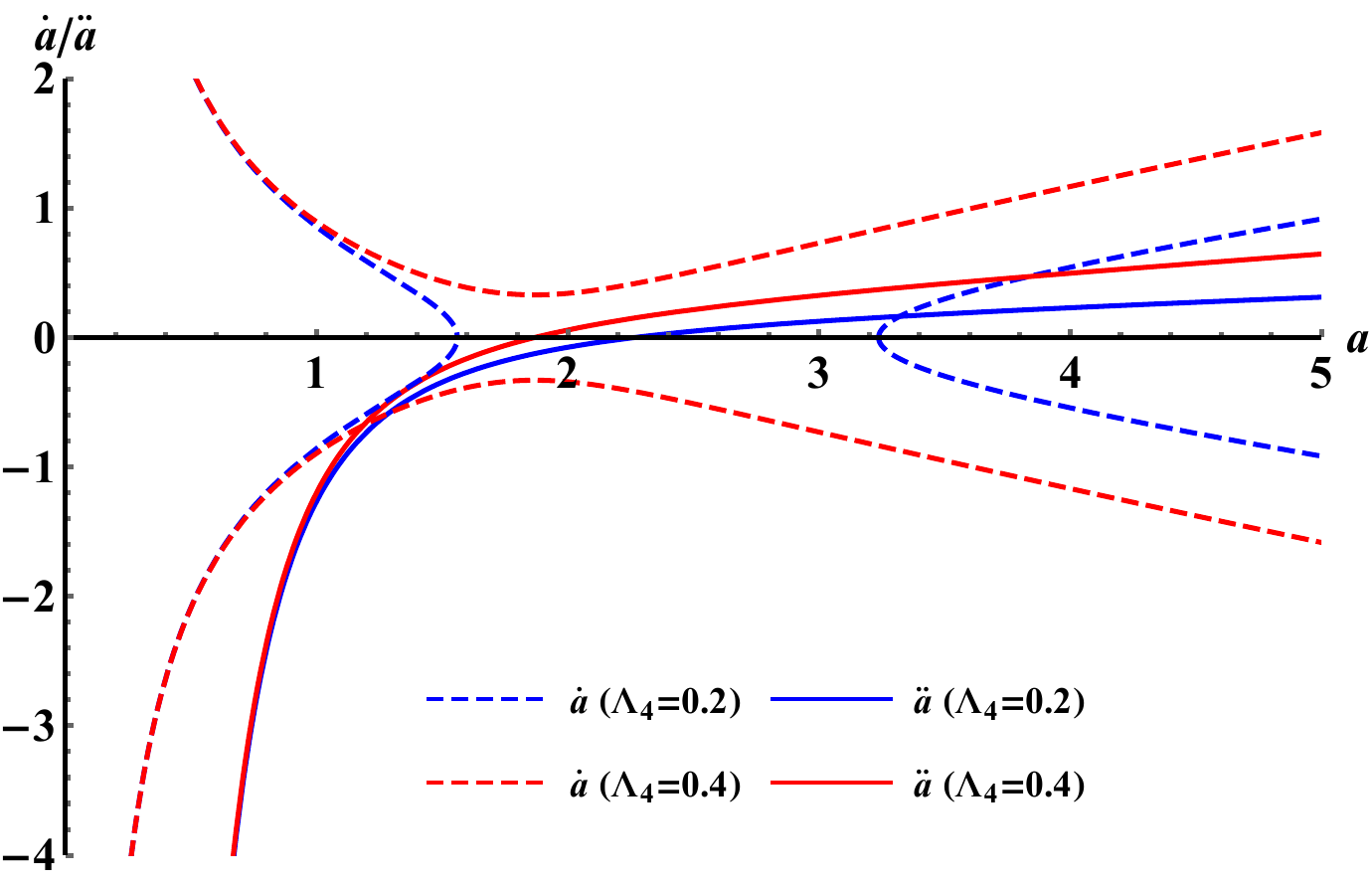}}
\hfill
\subfigure[$m_+=m_-=1$, $\Lambda_4=0.2$, $l_+=1.5$ and $l_-=1$]{\includegraphics[width=.45\linewidth]{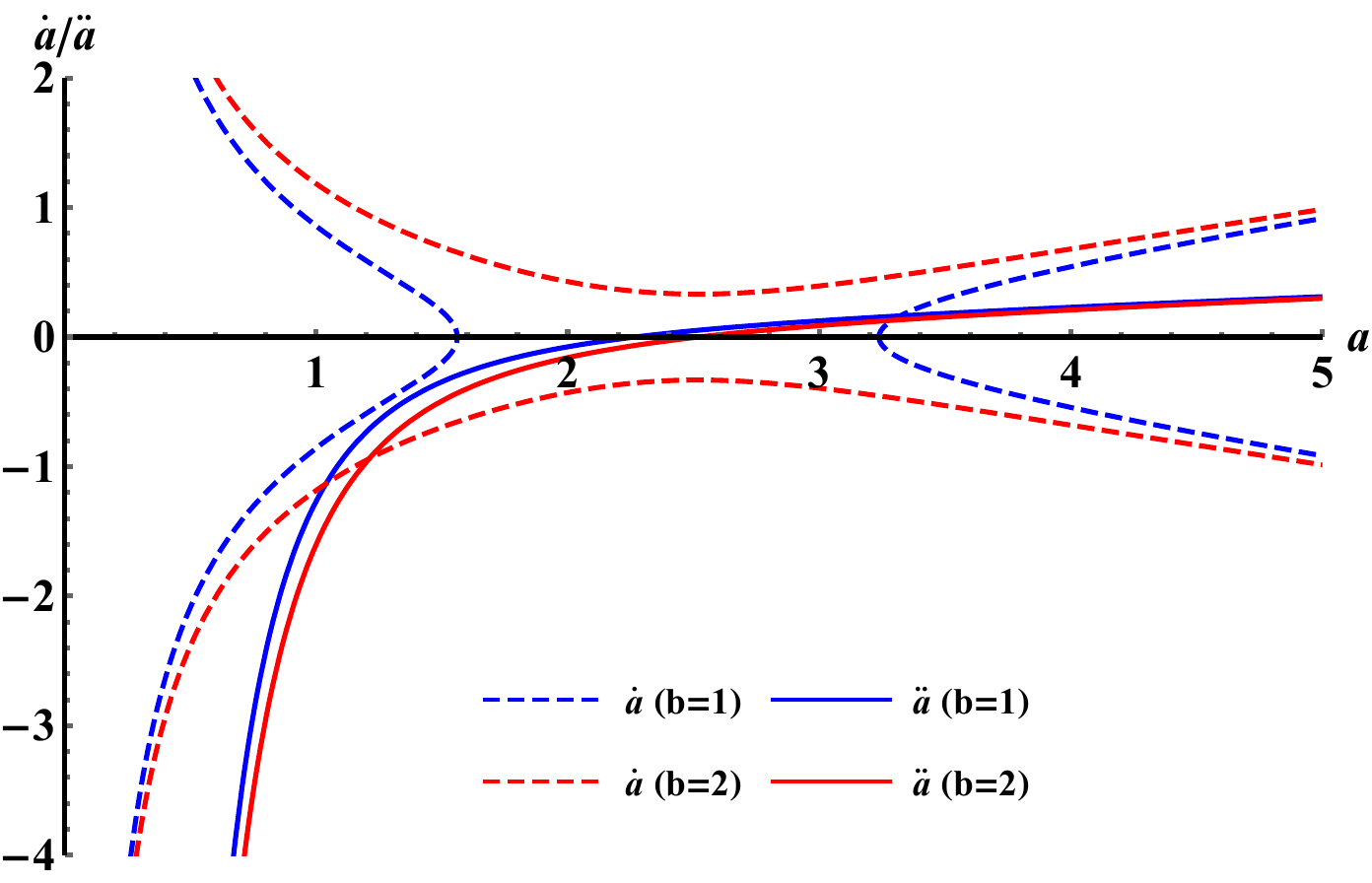}} \\[0.5cm]

\subfigure[ $m_+=0.5$, $m_-=1$, $b_+=b_-=3$, $l_+=1.5$ and $l_-=1$]{\includegraphics[width=.45\linewidth]{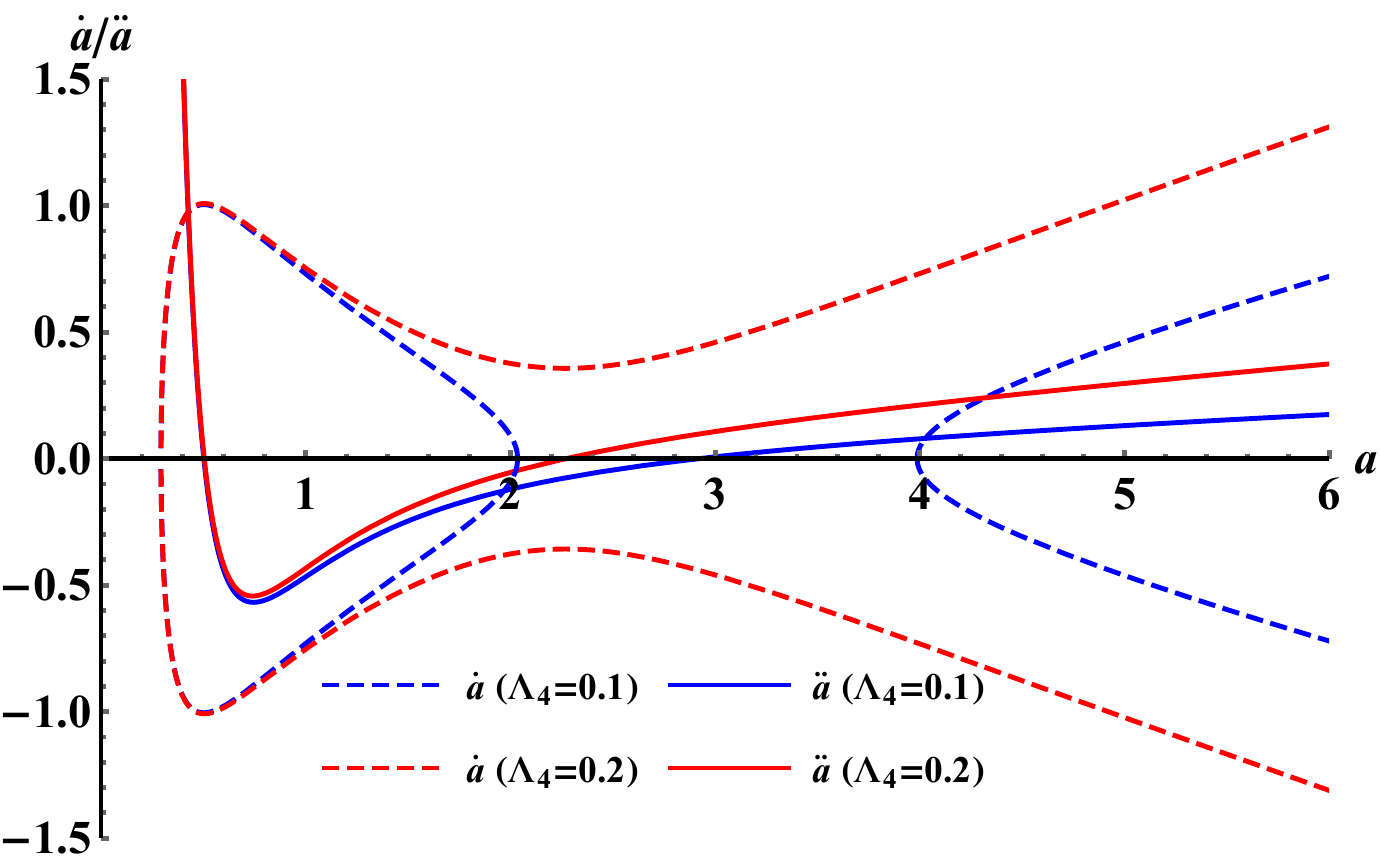}}
\hfill
\subfigure[ $m_+=0.5$, $m_-=1$, $\Lambda_4=0.1$, $l_+=1.5$ and $l_-=1$]{\includegraphics[width=.45\linewidth]{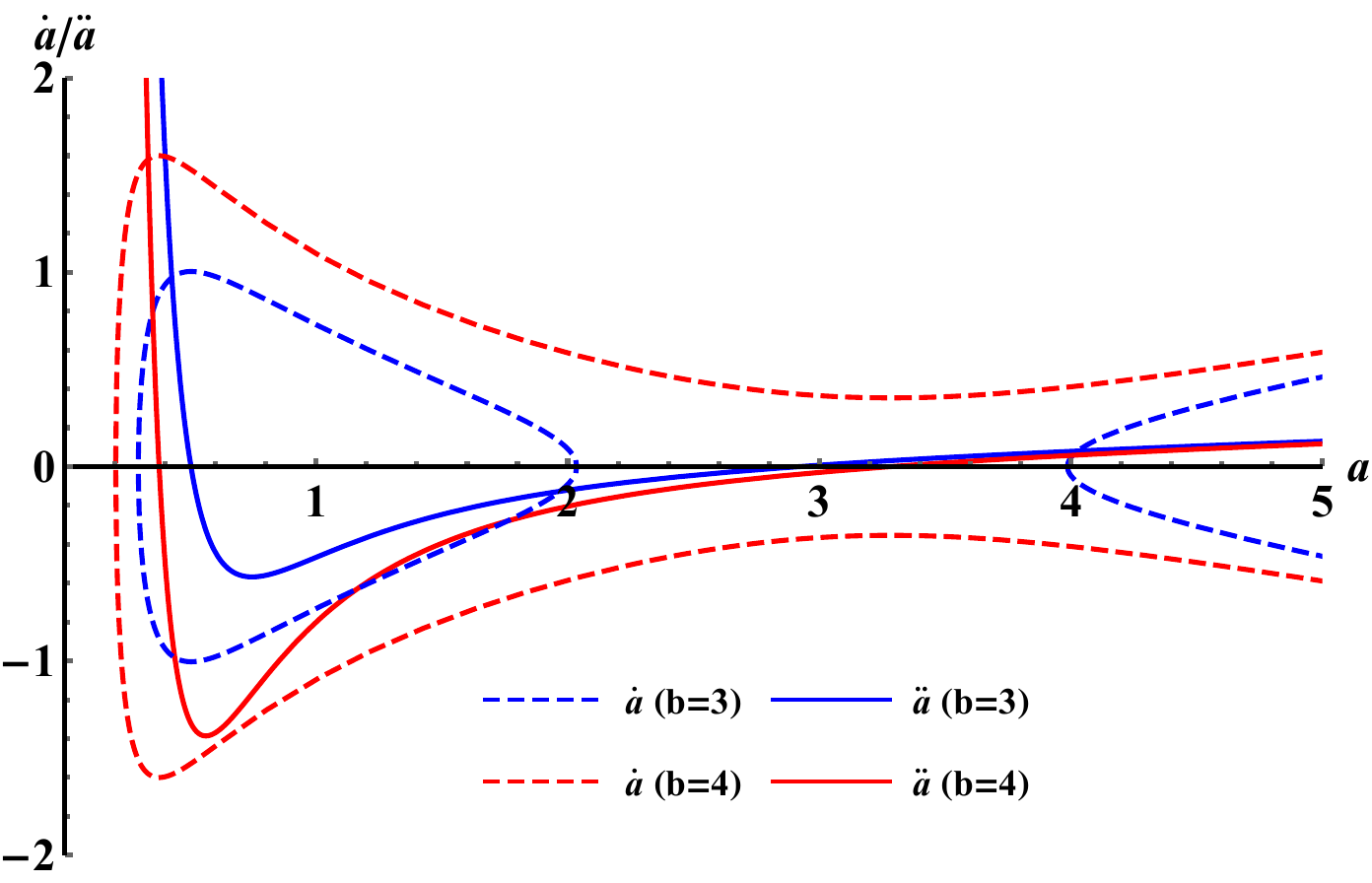}}

\caption{Plot of $\dot{a}$ and $\ddot{a}$ against $a$ for different values of $\Lambda_4$ and string density. (a) and (b) correspond to the condition $m_+l_+>m_-l_-$ and (c) and (d) correspond to the condition $m_+l_+<m_-l_-$. }
\label{fig 12}
\end{figure}

\begin{figure}[h]
\centering
\includegraphics[width=.5\linewidth]{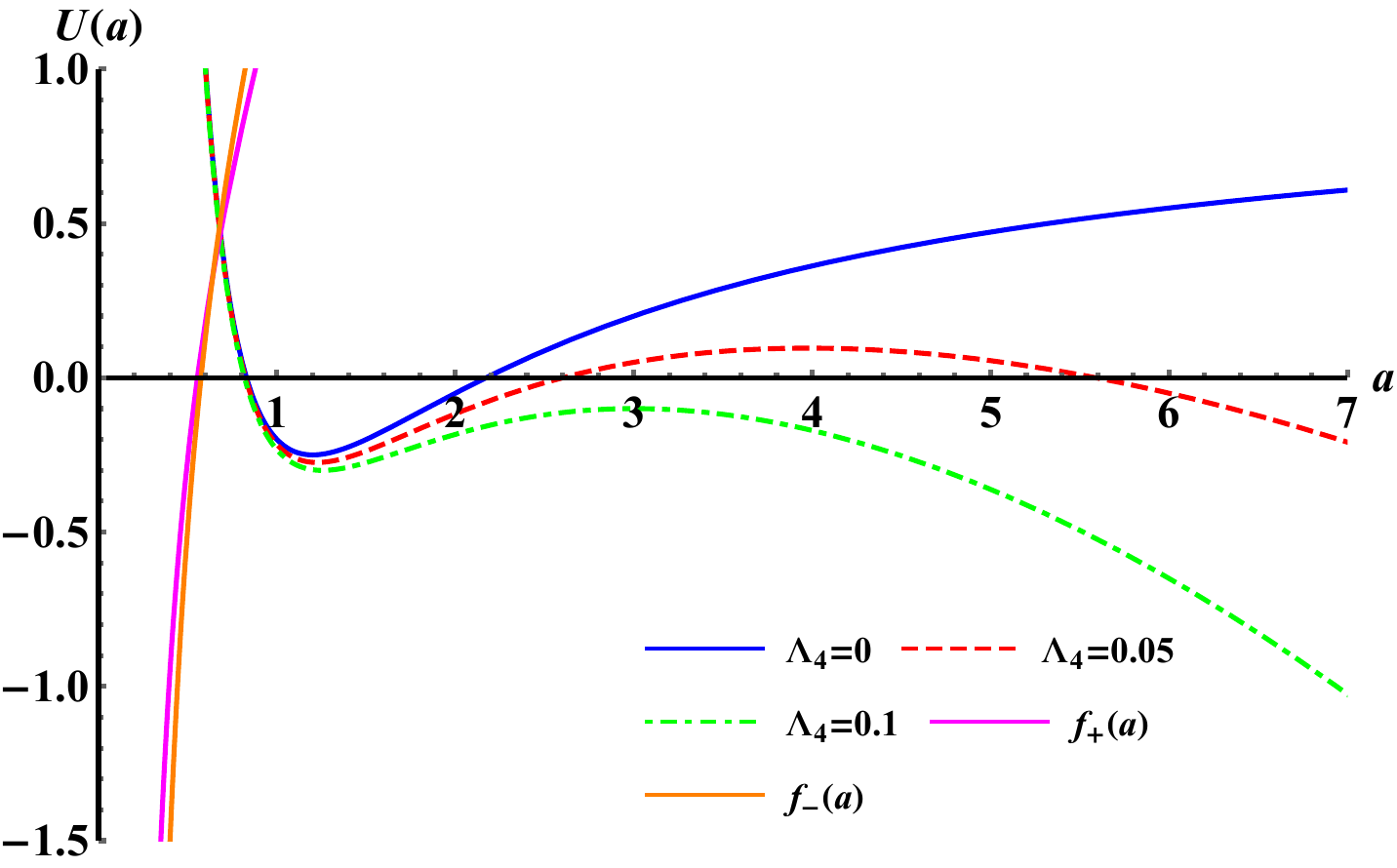}
\caption{Plot of $U(a)$ and $f(a)$ against $a$ for different values of $\Lambda_4$ where the bounce occurs outside the horizon of the black hole. The values of parameters are $m_+=0.2$, $m_-=0.4$, $b_+=0.5$, $b_-=0.1$, $l_+=1.1$ and $l_-=1$.}
\label{fig 13}
\end{figure}

However, in order to obtain a viable bounce in the shellworld scenario, we need the bounce to occur outside the horizon of the black hole. The condition for a viable bounce should be $a_b>a_h$, where $a_b$ is the size of the shellworld universe at the bounce and $a_h$ is the size of the horizon of the black hole. The solution of $f_{\pm}(a_h)=0$ gives the horizon of the black hole. In our case, the analytical solution is not  possible to derive. We thus resort to a parametric analysis to find a feasible bounce in the shellworld scenario. Under certain tuning of the parameters, we find that a viable bounce can be obtained for the condition $m_+l_+<m_-l_-$ and $b_+l_+>b_-l_-$, where the bounce is induced by the negative energy density of the dark radiation term. Figure (\ref{fig 13}) shows the plot of $U(a)$ and $f(a)$ against $a$ for different values of $\Lambda_4$ where the bounce occurs outside the horizon of the black hole. 

\subsection*{(II) $m_{\pm}<0$:}

We now discuss the effect of a negative mass parameter in the shellworld scenario. In this case, the Friedmann equation (\ref{FRW_eqn_shell}) is derived as,
\begin{equation}
\frac{\dot{a}^2}{a^2}=-\frac{1}{a^2}-\frac{1}{a^4}\left( \frac{ m_+l_+-m_-l_-}{l_+-l_-}\right)+\frac{2}{3a^3}\left( \frac{b_+l_+ -b_-l_- }{l_+-l_-}\right) +\frac{8\pi G_4}{3}\Lambda_4.
\end{equation}
For $\Lambda_4=0$, the solution of the Friedmann equation in conformal time $\eta$ is given by,
\begin{equation}\label{bubble_a_nm}
a(\eta)=\sqrt{\frac{1}{9}\left( \frac{b_+l_+ -b_-l_- }{l_+-l_-}\right)^2-\left( \frac{ m_+l_+-m_-l_-}{l_+-l_-}\right)}\sin \eta +\frac{1}{3}\left( \frac{b_+l_+ -b_-l_- }{l_+-l_-}\right).
\end{equation}
The conditions for a bouncing and cyclic nature to occur are $ (b_+l_+-b_-l_-)^2\geq 9(m_+l_+-m_-l_-)(l_+-l_-)$, $ m_+l_+> m_-l_- $ and $ b_+l_+ > b_-l_- $.
The second Friedmann equation in this case is given by,
\begin{equation}	
\frac{\ddot{a}}{a}=\frac{1}{a^4}\left( \frac{ m_+l_+-m_-l_-}{l_+-l_-}\right)-\frac{1}{3a^3}\left( \frac{b_+l_+ -b_-l_- }{l_+-l_-}\right) +\frac{8\pi G_4}{3}\Lambda_4.
\end{equation}

\begin{figure}[H]
\centering
\subfigure{\includegraphics[width=0.45\linewidth]{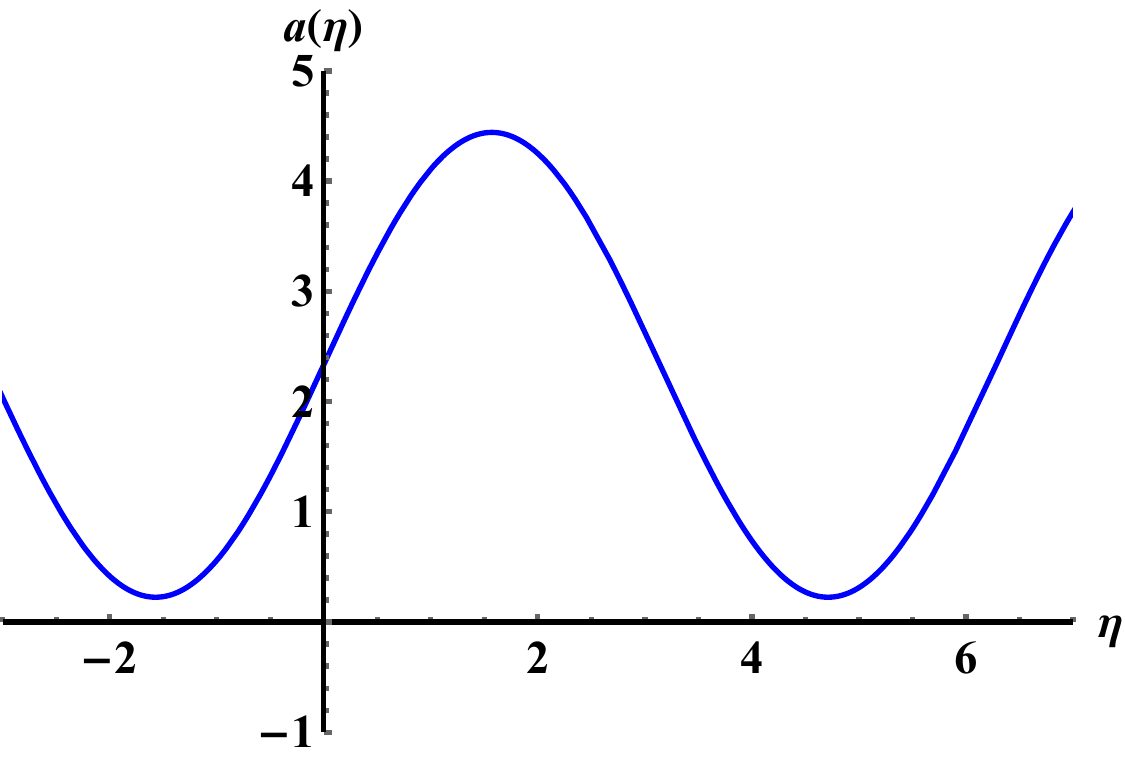}}
\subfigure{\includegraphics[width=0.45\linewidth]{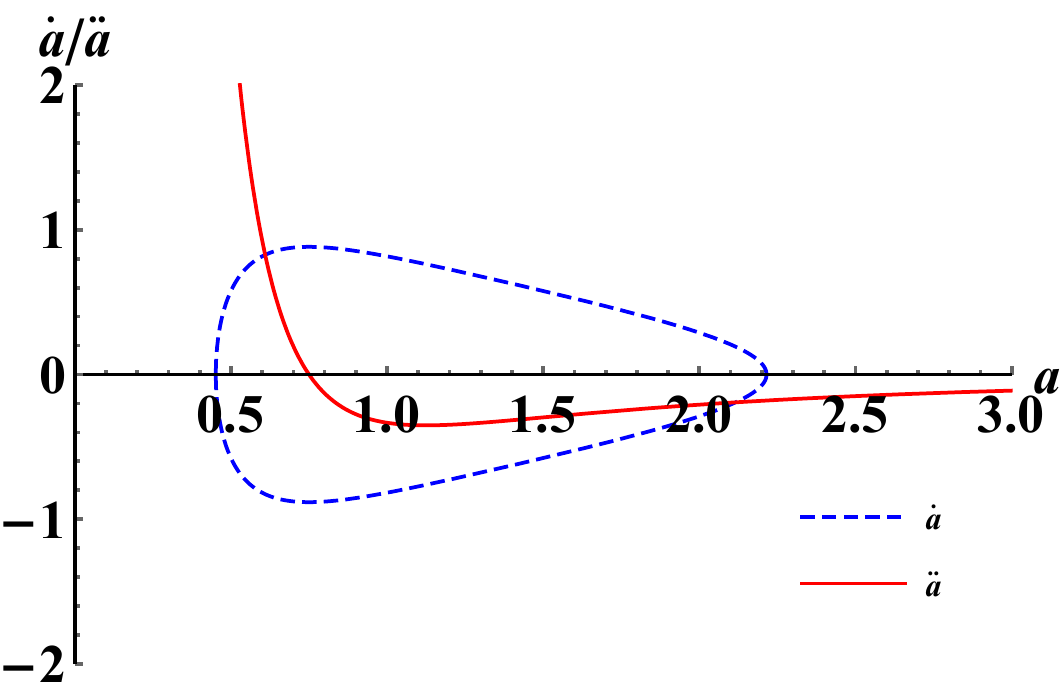}}	
\caption{(a) Plot of $a$ with respect to $\eta$. (b) Plot of $\dot a$ and $\ddot a$ with respect to $a$ for the non-singular universe.
The values of parameters are $m_+=1$, $m_-=1$, $b_+=2$, $b_-=1$, $l_+=1.5$ and $l_-=1$. }
\label{fig 14}
\end{figure}

Figure (\ref{fig 14}a) shows the plot of $a$ as a function of $\eta$ indicating the nonsingular bouncing and cyclic nature of the shellworld universe, while figure (\ref{fig 14}b) shows the plot of $\dot{a}$ and $\ddot{a}$ with respect to the size of the shellworld universe verifying the existence of nonsingular bouncing and cyclic nature of the scale factor.

\begin{figure}[H]
\centering	
\subfigure[$m_+=m_-=0.5$, $b_+=b_-=3$, $l_+=1.5$ and $l_-=1$]{\includegraphics[width=.45\linewidth]{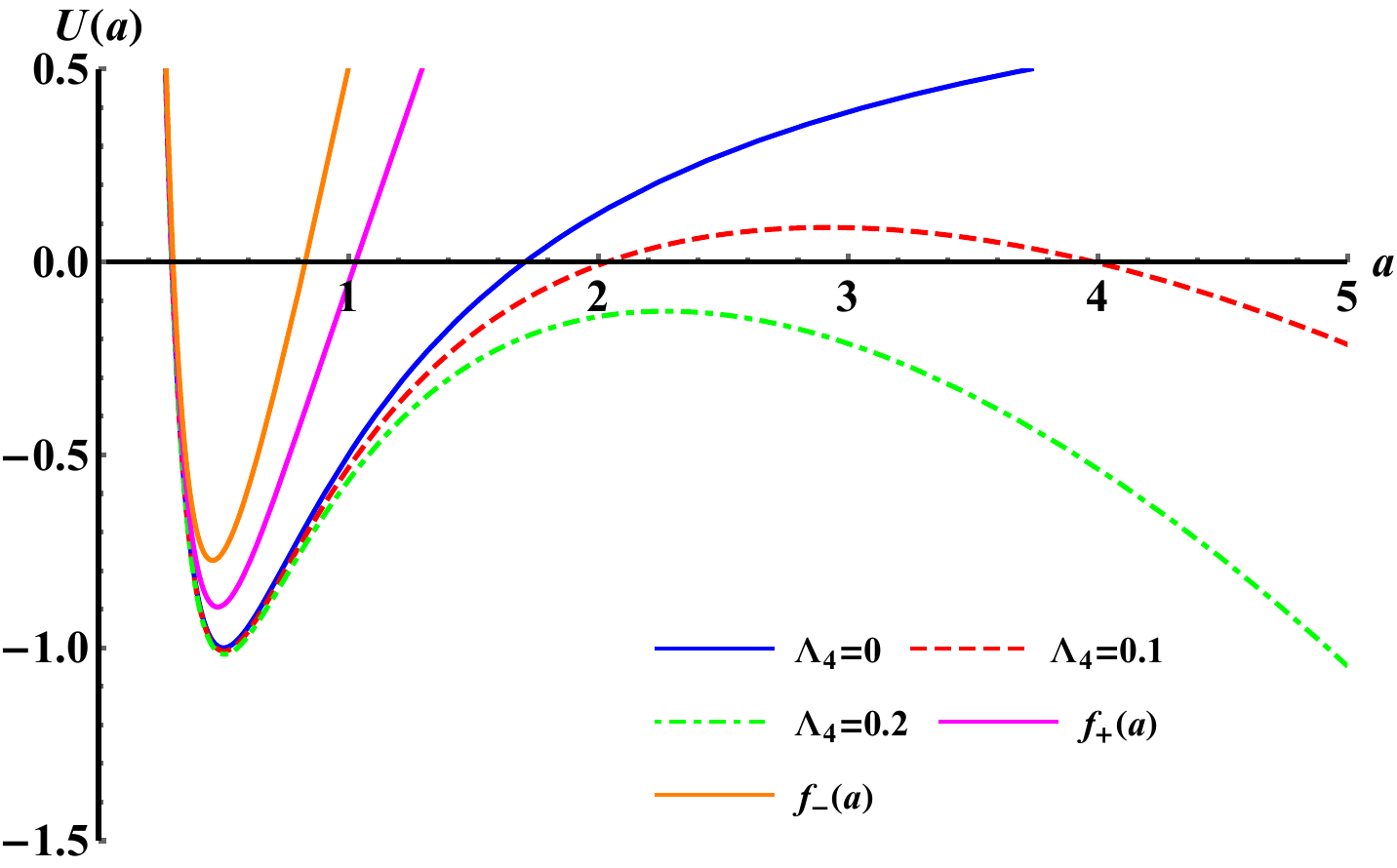}}
\hspace{0.2cm}
\subfigure[$m_+=m_-=0.5$, $\Lambda_4=0.1$, $l_+=1.5$ and $l_-=1$]{\includegraphics[width=.45\linewidth]{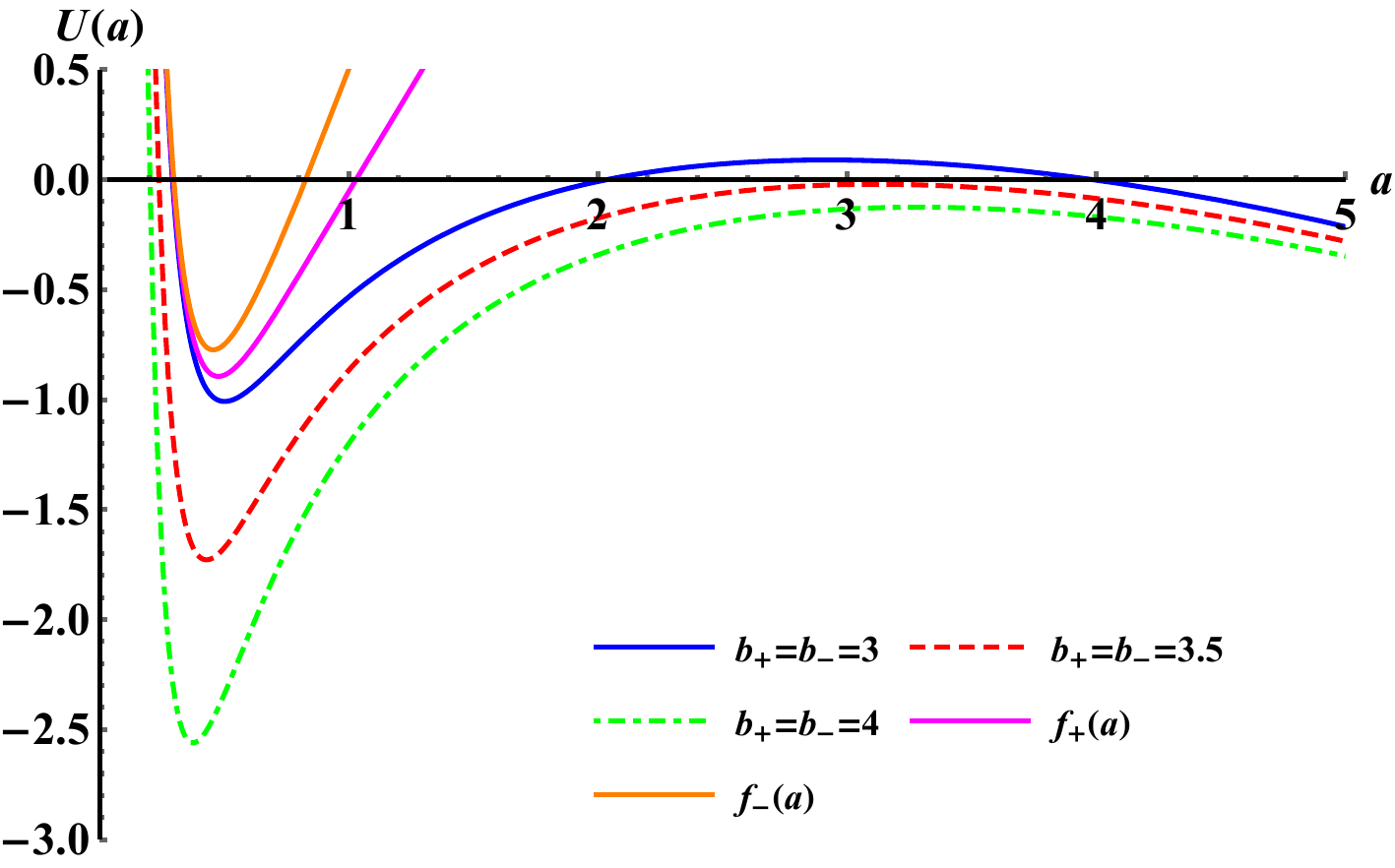}}
\caption{Plot of $U(a)$ vs $a$ for different values of (a) $\Lambda_4$ and (b) string density.}
\label{fig 15}	
\end{figure}

\begin{figure}[h]
\centering
\includegraphics[width=.5\linewidth]{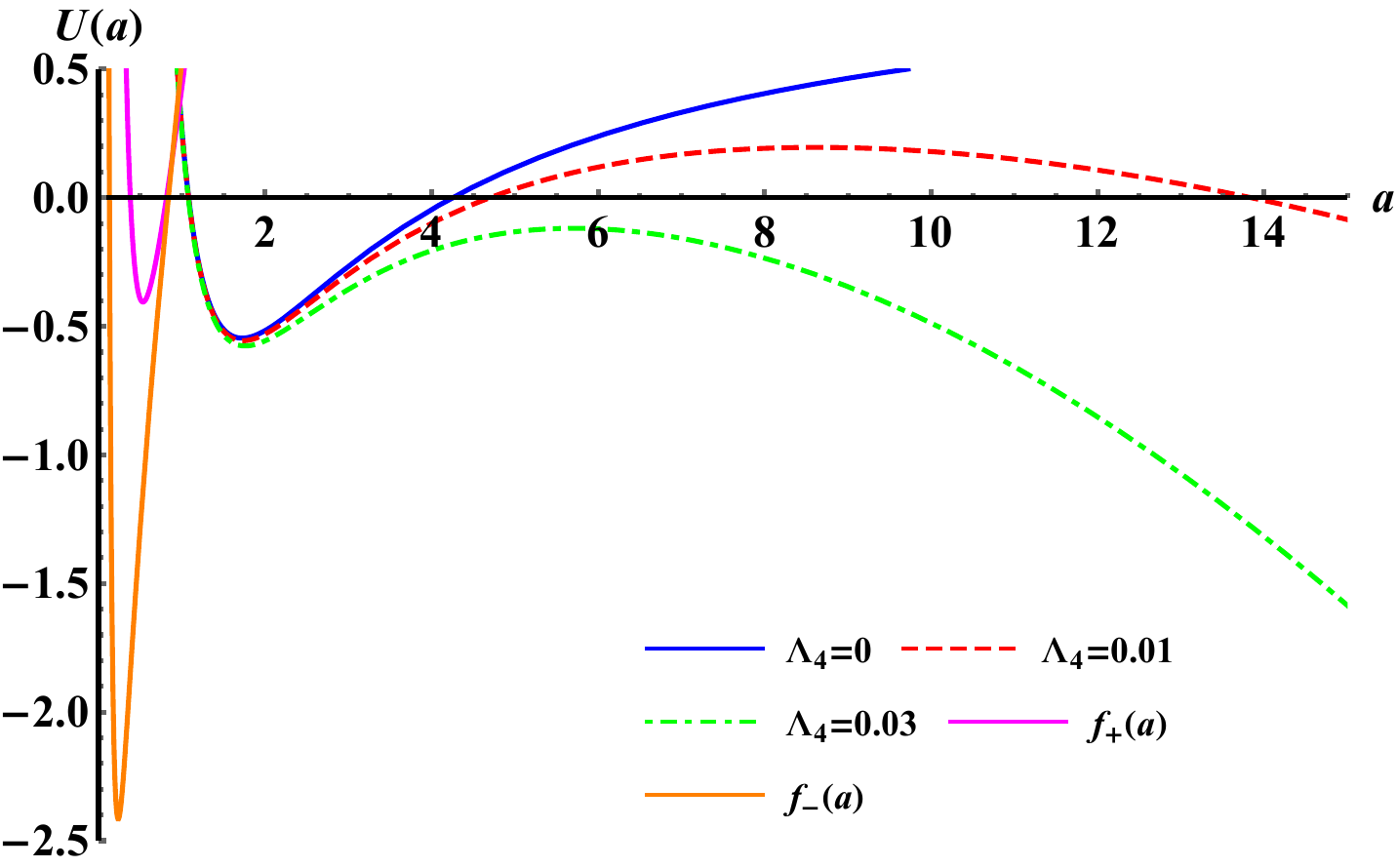}
\caption{Plot of $U(a)$ and $f(a)$ against $a$ for different values of $\Lambda_4$ where the bounce occurs outside the horizon of the black hole. The values of parameters are $m_+=0.6$, $m_-=0.2$, $b_+=3$, $b_-=2.5$, $l_+=1.1$ and $l_-=1$.}
\label{fig 16}
\end{figure}

\begin{figure}[H]
\centering	
\subfigure[$m_+=m_-=0.5$, $b_+=b_-=3$, $l_+=1.5$ and $l_-=1$]{\includegraphics[width=.45\linewidth]{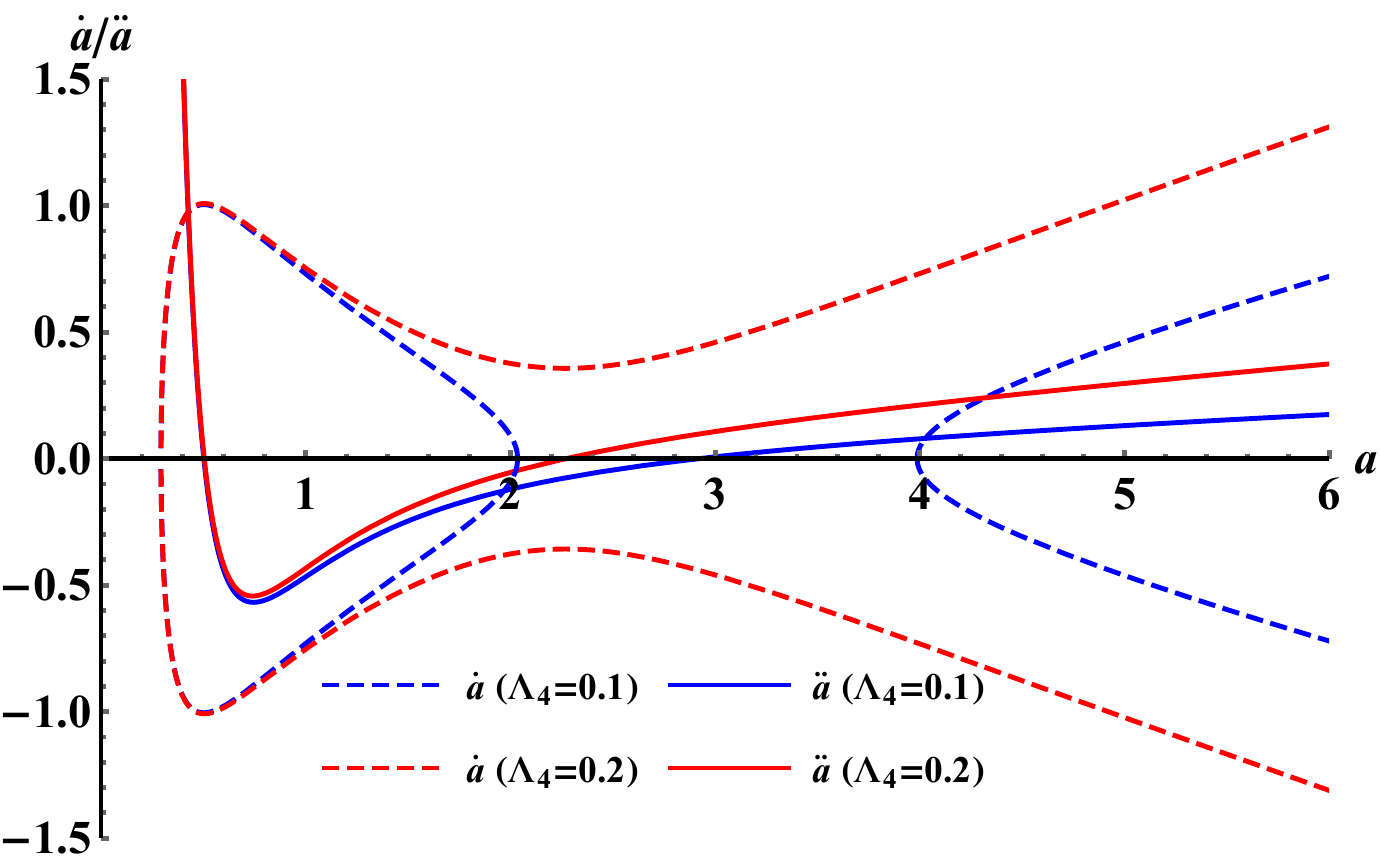}}
\hspace{0.2cm}
\subfigure[$m_+=m_-=0.5$, $\Lambda_4=0.1$, $l_+=1.5$ and $l_-=1$]{\includegraphics[width=.45\linewidth]{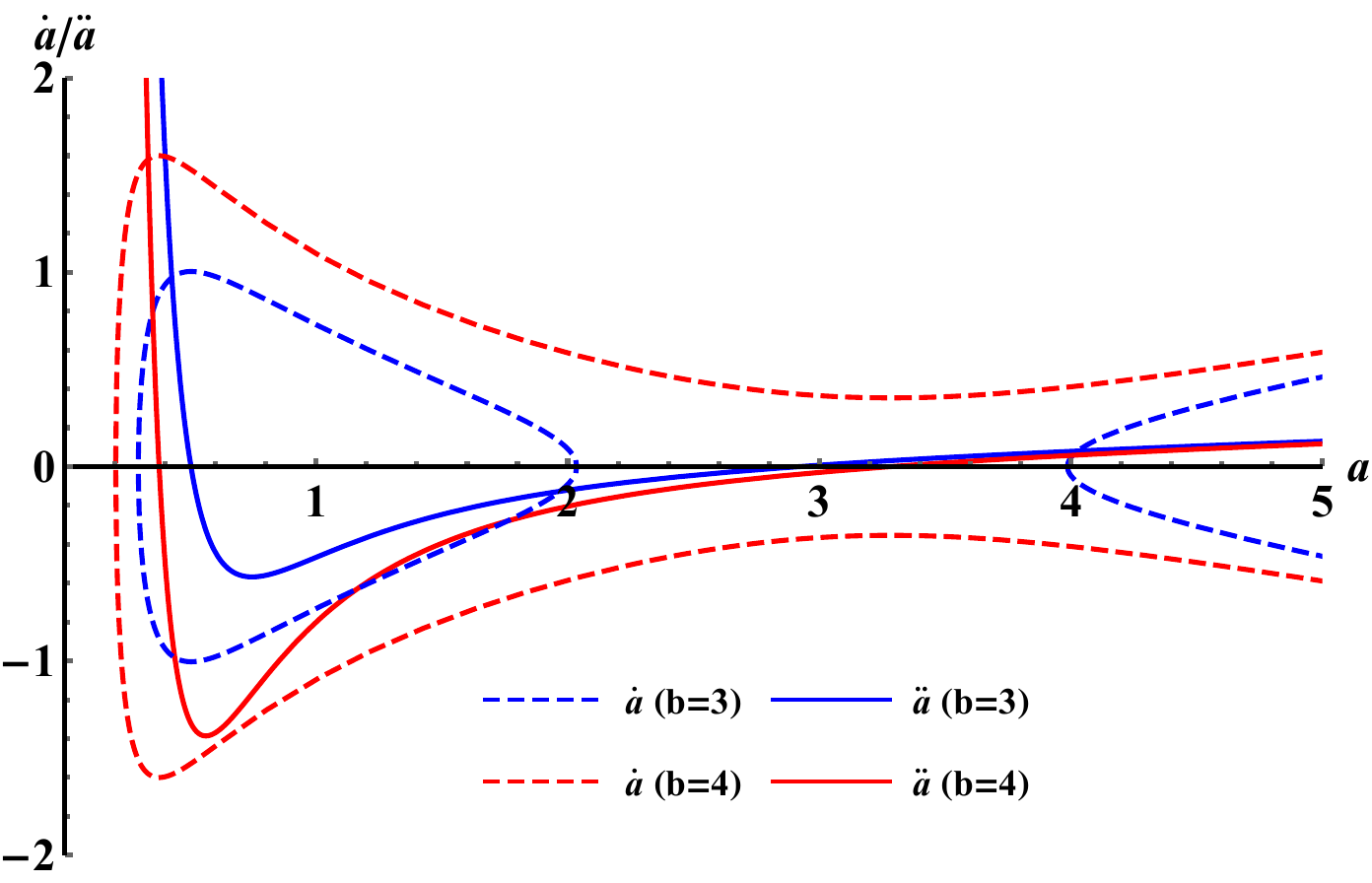}}
\caption{Plot of $\dot{a}$ and $\ddot{a}$ vs $a$ for different values of (a) $\Lambda_4$ and (b) string density.}
\label{fig 17}	
\end{figure}

The effective potential of the shellworld universe in this case is given by,
\begin{equation}
U(a)=1+\frac{1}{a^2}\left( \frac{ m_+l_+-m_-l_-}{l_+-l_-}\right)-\frac{2}{3a}\left( \frac{b_+l_+ -b_-l_- }{l_+-l_-}\right)-\frac{8\pi G_4}{3}\Lambda_4 a^2.
\end{equation}
Here, as in the previous case, we have plotted $U(a)$ against $a$ for different values of $\Lambda_4$ and string density in figure (\ref{fig 15}). Two regions of the shellworld universe are observed where we can obtain a nonsingular bouncing nature. One is the bounce induced by the spatial curvature term which is outside the horizon of the black hole and the other is the bounce induced by the negative energy density of the dark radiation term which is inside the horizon of the black hole. 
However, by careful tuning of the parameters, we find that a viable bounce can be obtained in the region outside the black hole horizons which is shown in figure (\ref{fig 16}).
The plot of $\dot{a}$ and $\ddot{a}$ against $a$ in figure (\ref{fig 17}) for different values of $\Lambda_4$ and string density verifies the two regions of nonsingular bouncing nature of the shellworld universe as discussed above. Above certain values of brane cosmological constant and string density, the two regions merge and only one region of universe exists where the universe begins from a nonsingular value of scale factor and expands eternally. 

\subsection*{(III) $b_{\pm} = 0$:}

In absence of cloud of strings $(b_{\pm}=0)$, the Friedmann equation of the shellworld universe can be written as,
\begin{equation}
\frac{\dot{a}^2}{a^2}=-\frac{1}{a^2}+\frac{1}{a^4}\left( \frac{ m_+l_+-m_-l_-}{l_+-l_-}\right)+\frac{8\pi G_4}{3}\Lambda_4.
\end{equation}
The solution of the above equation in proper time $\tau$ is given by,
\begin{equation}
a(\tau)=\frac{1}{\sqrt{2\lambda}}[1+\sqrt{4\lambda M-1}\sinh(2\sqrt{\lambda}\tau)]^{1/2},
	\end{equation}
	where, $\lambda=\frac{8\pi G_4}{3}\Lambda_4$ and $M=\left( \frac{ m_+l_+-m_-l_-}{l_+-l_-}\right)$. The solution is valid only when $4\lambda M\geq 1$ and the scale factor begins from a non-zero minimum value of $a_{min}=\frac{1}{\sqrt{2\lambda}}$ at $\tau=0$ and expands eternally with the bounce induced by the curvature term. Such a scenario is discussed in \cite{Biswas:2003rb,Petkou2002} in detail for the braneworld framework.

\section{Fluctuations of brane/bubble fields} \label{sec 4}

In the previous section, we have discussed the conditions under which a nonsingular bouncing and cyclic nature of the universe can be obtained in the braneworld and shellworld scenarios. We have also shown that the bounce can occur outside the horizon of the black hole under certain tuning of the parameters. However, the stability of perturbations in bouncing cosmological models remains an open issue. While a complete treatment would involve analyzing perturbations in the full five-dimensional background, the implementation of such a framework in the present setup is highly nontrivial and remains unresolved. Nevertheless, one can still investigate the stability of the bouncing solution within an effective framework. We analyze the fluctuations of a massless minimally coupled scalar field $\phi$ localized on the brane, treating it as a test field whose contribution to the background cosmological evolution can be neglected. \\
We consider the Fourier component $\delta \phi_k$ of the perturbation with comoving wave number 
$k$, and define the Mukhanov variable $v_k \equiv a \delta \phi_k$. The corresponding Mukhanov Sasaki \cite{Mukhanov:1985rz,Sasaki:1986hm} equation turns out to be,
\begin{equation}\label{MS_eqn}
	v_k^{''} + \left(k^2-\frac{a^{''}}{a}\right)v_k=0.
\end{equation}
Here, $(')$ denotes differentiation with respect to the conformal time $\eta$. The solution of the Friedmann equation (\ref{FRW_eqn},\ref{FRW_eqn_shell}) in conformal time $\eta$ which shows a nonsingular bouncing and cyclic nature of the brane/bubble universe is given by,
\begin{equation}
	a(\eta)=B + \sqrt{B^2-M} \sin\eta,
\end{equation}
where, for the braneworld scenario, $B=\frac{b}{3}$ and $M=m$, while for the shellworld scenario,
\begin{equation}
	B=\frac{1}{3}\left( \frac{b_+l_+ - b_-l_-}{l_+-l_-}\right)  \hspace{0.2 cm} \text{and}  \hspace{0.2 cm} M=\left( \frac{m_+l_+ - m_-l_-}{l_+-l_-}\right) .
\end{equation}
The term $a''/a$ in equation (\ref{MS_eqn}) at the bounce $(\sin \eta=-1)$ equates as,
\begin{equation}
	\frac{a''}{a}=-\frac{\sqrt{B^2-M}\sin \eta}{B+\sqrt{B^2-M}\sin \eta }=\frac{\sqrt{B^2-M}}{B-\sqrt{B^2-M}}.
\end{equation}
Since, $B>\sqrt{B^2-M}$, this ensures ${a''}/{a}$ remains finite.

In order to analyze the evolution of the perturbation mode $\delta \phi_k$, we consider two regions corresponding to $\eta<\eta_o$ which represents the short wave limit ($k^2 \gg a''/a$) and $\eta>\eta_o$ which represents the long wave limit ($k^2 \ll a''/a$). At the time $\eta_o$ the terms $k$ and $a''/a$ in the equation (\ref{MS_eqn}) becomes equal, leading to the condition,
\begin{equation}
	k^2=\frac{a''}{a}=-\frac{\sqrt{B^2-M}\sin \eta_o}{B+\sqrt{B^2-M}\sin \eta_o}.
\end{equation}

In the short wave limit ($k^2 \gg a''/a$), the solution of equation (\ref{MS_eqn}) for the canonically normalized field $v_k$ is given by,

\begin{equation}\label{short_wave}
	v_k\equiv \frac{e^{-i k \eta}}{\sqrt{2k}} \Rightarrow
	\delta \phi_k = \frac{v_k}{a} \equiv \frac{e^{-i k \eta}}{\sqrt{2k}a},
\end{equation}
while, in the long wave limit ($k^2\ll a''/a$), it yields,
\begin{equation}\label{long_wave}
	\delta \phi_k = \frac{v_k}{a} \equiv A_1 + A_2 \int \frac{d\eta}{a^2}.
\end{equation}
Here, $A_1$ and $A_2$ are constants. The integral $\int d\eta/a^2$ is computed as,
\begin{align}
	\int \frac{d\eta}{a^2} = & \int \frac{d\eta}{\left(\sqrt{B^2-M} \sin\eta + B\right)^2} \\  \nonumber = & \frac{2 B \tan ^{-1}\left(\frac{\sqrt{B^2-M}+B \tan \left(\frac{\eta}{2}\right)}{\sqrt{M}}\right)}{M^{3/2}}+\frac{2 \cos (\eta) \left(\left(M-B^2\right) \sin (\eta)+B \sqrt{B^2-M}\right)}{M \left(\left(B^2-M\right) \cos (2 \eta)+B^2+M\right)}.
\end{align}

The integration constants $A_1$ and $A_2$ of equation \ref{long_wave} are estimated by imposing the continuity of both the perturbation mode $(\delta\phi_k)$ and its first derivative across the matching surface at $(\eta=\eta_o)$. Finally, the integration constant $A_1$ and $A_2$ are found to be,
\begin{equation}
|A_1|= \frac{1}{\sqrt{2k}}\sqrt{\left( \frac{1+k^2}{B} + \frac{P}{1+k^2}Q\right)^2 + \left(\frac{Bk}{1+k^2}Q \right)^2}
\end{equation}
where,
\begin{align}
P=&\sqrt{B^2(1+2k^2)-M(1+k^2)^2}\\
Q=&\frac{P}{BM} + \frac{2B}{M^{3/2}}\arctan \left[{-\sqrt{\frac{M}{B^2-M}}+\frac{B^2}{\sqrt{M(B^2-M)}}\left( 1-\frac{k^2\sqrt{B^2-M}}{(1+k^2)\sqrt{B^2-M}+P} \right)}\right],
\end{align}
and
\begin{equation}
|A_2|= \frac{1}{\sqrt{2k}}\sqrt{B^2\left( \frac{1+3k^2}{(1+k^2)^2}\right)-M }.
\end{equation}
The above analysis indicates the scalar field mode functions remain regular throughout the cosmological evolution, and the corresponding fluctuations do not develop any divergences as the universe passes through the bounce. This indicates that the nonsingular background does not induce any pathological behavior in the evolution of the test scalar field, providing evidence that the bounce is stable against linear scalar field fluctuations.

It is important to emphasize, however, that the present analysis is performed in the test-field approximation, where the scalar field is assumed to have a negligible effect on the background geometry. A definitive assessment of the stability of the bouncing solution would require a fully coupled perturbation analysis of both the brane and the bulk geometry. In the present model, where the bulk consists of an AdS Schwarzschild spacetime containing a cloud of strings, such an analysis is highly nontrivial. One must consistently account for perturbations of the higher-dimensional gravitational field, the string cloud, and their coupling to the brane through the junction conditions. This represents a substantially more challenging problem and lies beyond the scope of the present work. Nevertheless, the finiteness of the scalar field fluctuations across the bounce provides an important consistency check of the background solution in the shellworld scenario.
\section{Summary} \label{sec 5}
In this work, we have studied the cosmological evolution of the four dimensional brane/bubble universe in an AdS Schwarzschild bulk spacetime with the incorporation of a cloud of strings. The string cloud is uniformly distributed and stretches along the radial direction of the bulk, with its endpoints attached to the four-dimensional brane placed perpendicular to the radial coordinate of the bulk.
 The attached end points of the strings represent massive particles or matter on the four dimensional brane. The cloud of strings deform the AdS bulk spacetime and allows the mass parameter of the bulk to take negative values at smaller size of the bulk spacetime on increasing values of the string density. We have derived the effective Friedmann equations on the brane in which the mass parameter term gives a dark radiation contribution and the string density term gives a matter contribution.

In the braneworld framework where the two bulk regions on either side of the brane are identical and related by a $\mathbb{Z}_2$ symmetry, we have investigated the cosmological evolution of the brane universe using the Friedmann equations for both critical and non-critical branes. The exact analytical solutions of the Friedmann equation is derived for the critical brane. In the case of a non-critical brane, we have performed a qualitative analysis of the nature of cosmological evolution by treating the evolution equation of the brane universe as the dynamics of a classical particle moving in an effective potential.

The critical brane universe with a positive mass parameter shows the brane universe to begin from a nonsingular value of scale factor and collapse back to singularity for the closed geometry while for open and flat geometries, the brane universe expands exponentially and quadratically respectively from a singularity. 
For a non-critical brane with a positive mass parameter, there are two possible regions for the universe to exist within a certain range of brane cosmological constant and below a certain critical value of string density for a closed universe with a positive brane cosmological constant. In one region, the universe undergoes a nonsingular bounce induced by the spatial curvature term outside the horizon of the black hole. Whereas in the other region the universe crunches from an expanding mode to a singularity. Above the certain range of brane cosmological constant and critical value of string density, the universe does not have any turning point and expands or crunches continuously. For a negative brane cosmological constant, the universe quickly turns back from expanding mode to contracting mode and finally collapses to the singularity. For an open and flat universe with a positive brane cosmological constant, the universe eternally expands without a bouncing nature while for a negative brane cosmological constant, the universe goes from an expanding nature to a contracting nature and finally collapses to the singularity.

In the case of a negative mass parameter, the brane universe has a bouncing and cyclic nonsingular nature for a closed critical brane universe. Whereas, for an open and flat critical brane universe, the scale factor undergoes a nonsingular exponential and quadratic expansion respectively. For a closed universe with a positive brane cosmological constant, again there are two possible regions for the universe to exist depending on $\Lambda_4$ and string density. The region with larger scale factor shows a nonsingular bounce where the bounce is induced by the spatial curvature term and occurs outside the horizon of the black hole while the region with smaller scale factor shows a nonsingular bounce where the bounce is induced by the negative energy density of the dark radiation term and occurs inside the Cauchy horizon of the black hole.
Above the certain value of $\Lambda_4$ and string density, two regions merge and only one region of universe exists where the universe begins from a nonsingular value of scale factor and expands eternally. For open and flat universe with a positive brane cosmological constant, the universe gradually expands from a nonsingular value of scale factor. For a negative brane cosmological constant, the universe has a nonsingular bouncing and cyclic nature for all three geometries of the brane. However, the bounce obtained from the negative energy density of the dark radiation term occurs inside the Cauchy horizon of the black hole and thus suffers from the Cauchy horizon instability. 

Consequently, we further studied the bouncing scenario in the shellworld model to explore stable nonsingular bouncing cosmological model. The shellworld or the dark bubble scenario lacks $\mathbb{Z}_2 $ symmetry with two distinct AdS vacua representing an inside and outside region. We have derived the Friedmann equations on the shellworld in the presence of cloud of strings for a positive mass parameter. We then obtain an exact solution of the Friedmann equation for the critical shellworld universe. The geometry of the shellworld allowed the dark radiation term to contribute a negative energy density even in the case of a positive mass parameter. This negative energy density contribution from the dark radiation term induces a nonsingular bouncing and cyclic nature of the shellworld universe. A qualitative analysis of the nature of cosmological evolution for a non-critical shellworld universe is also performed. As in the braneworld case, in this configuration we observe two regions of the shellworld universe where we can obtain a nonsingular bouncing nature for a positive brane cosmological constant. One is the bounce induced by the spatial curvature term which occurs at higher values of the scale factor within a certain range of $\Lambda_4$ and below a critical value of $b$ outside the horizon of the black hole, and the other is the bounce induced by the negative energy density of the dark radiation term which is inside the horizon of the black hole. But, under certain tuning of parameters, the bounce can be obtained outside the horizon of the black hole. As the value of $\Lambda_4$ and string density increases, the regions merge and only one region of universe exists where the universe begins from a nonsingular value of scale factor and expands eternally. 

Further, we have analyzed the case of a negative mass parameter in the shellworld scenario. We also obtain a nonsingular bouncing nature of the shellworld universe in this case. As studied in the previous case, a negative mass parameter leads to two horizons in the bulk, nevertheless, we find that under certain tuning of the values of the parameters, the bounce of the shellworld universe occurs outside the horizons of the bulk spacetimes avoiding the instability. 

We have also analyzed the stability of the bouncing solution by studying the evolution of a test scalar field on the braneworld/shellworld. We find that the scalar field fluctuations remain finite across the bounce, indicating that the nonsingular background does not induce any pathological behavior in the evolution of the test scalar field. This provides evidence that the bounce remains regular under linear scalar field fluctuations. This study shows that the shellworld (dark bubble) scenario represents an effective framework for constructing a nonsingular bouncing cosmology. 

\section{Acknowledgment}
K. P. Sherpa would like to thank the MoTA, Govt. of India for providing financial support through NFST. We also thank I. K. P. Chettri for carefully reading the draft and offering valuable comments.

\printbibliography
\end{document}